\documentclass[article,aps,twocolumn,showpacs,amsmath,amssymb,superscriptaddress,nofootinbib,floatfix]{revtex4-2}

\usepackage{graphicx}
\usepackage{dcolumn}
\usepackage{bm}
\usepackage[colorlinks=true,linkcolor=blue,citecolor=blue,urlcolor=blue]{hyperref}

\usepackage{array,booktabs,tabularx}
\usepackage[caption=false]{subfig}
\usepackage[USenglish]{babel}
\usepackage{braket}
\usepackage{xcolor}
\usepackage{amsfonts}
\usepackage{amssymb}
\usepackage[normalem]{ulem}
\usepackage{amsmath}

\begin{document}

\title{Symmetry-Selective Stabilization of Charge-Density Wave
in ScV$_6$Sn$_6$}

\author{A. Korshunov}
\thanks{These authors contributed equally to this work.}
\affiliation{Donostia International Physics Center (DIPC), Paseo Manuel de Lardizábal. 20018, San Sebastián, Spain}

\author{C.-Y. Lim}
\thanks{These authors contributed equally to this work.}
\affiliation{Donostia International Physics Center (DIPC), Paseo Manuel de Lardizábal. 20018, San Sebastián, Spain}

\author{J. Corral-Sertal}
\thanks{These authors contributed equally to this work.}
\affiliation{Departamento de Física Aplicada, Universidade de Santiago de Compostela, E-15782 Campus Sur s/n, Santiago de Compostela, Spain}
\affiliation{CiQUS, Centro Singular de Investigacion en Quimica Biolóxica e Materiais Moleculares, Departamento de Quimica-Fisica, Universidade de Santiago de Compostela, Santiago de Compostela, E-15782, Spain}

\author{G. Garbarino}
\affiliation{European Synchrotron Radiation Facility (ESRF), BP 220, F-38043 Grenoble Cedex, France}

\author{D. Chernyshov}
\affiliation{Swiss-Norwegian BeamLines at European Synchrotron Radiation Facility, BP 220, F-38043 Grenoble Cedex, France}

\author{A. Rajapitamahuni}
\affiliation{National Synchrotron Light Source II, Brookhaven National Laboratory, Upton, New York 11973, USA}
\affiliation{Department of Physics, SRM University - AP, Amaravati, Andhra Pradesh, 522502,India.}
\affiliation{Department of Physics and Astronomy, University of Nebraska-Lincoln, Nebraska 68588, USA.}

\author{C. Yi}
\affiliation{Max Planck Institute for Chemical Physics of Solids, D-01187 Dresden, Germany}

\author{ S. Roychowdhury}
\affiliation{Max Planck Institute for Chemical Physics of Solids, D-01187 Dresden, Germany}
\affiliation{Department of Chemistry, Indian Institute of Science Education and Research Bhopal, Bhopal-462 066, India}

\author{C. Shekhar}
\affiliation{Max Planck Institute for Chemical Physics of Solids, D-01187 Dresden, Germany}

\author{C. Felser}
\affiliation{Max Planck Institute for Chemical Physics of Solids, D-01187 Dresden, Germany}

\author{V. Pardo}
\affiliation{Departamento de Física Aplicada, Universidade de Santiago de Compostela, E-15782 Campus Sur s/n, Santiago de Compostela, Spain}
\affiliation{Instituto de Materiais iMATUS, Universidade de Santiago de Compostela, E-15782 Campus Sur s/n, Santiago de Compostela, Spain} 

\author{Ella M. Schmidt}
\affiliation{MARUM - Center for Marine Environmental Sciences, University of Bremen, 28359, Bremen, Germany}
\affiliation{Crystallography and Geomaterial Research, Faculty of Geosciences, University of Bremen, 28359, Bremen, Germany}
\affiliation{MAPEX Center for Materials and Processes, University of Bremen, 28359, Bremen, Germany}

\author{S. Blanco-Canosa}
\email{sblanco@dipc.org}
\affiliation{Donostia International Physics Center (DIPC), Paseo Manuel de Lardizábal. 20018, San Sebastián, Spain}
\affiliation{IKERBASQUE, Basque Foundation for Science, 48013 Bilbao, Spain}

\date{\today}

\begin{abstract}
Charge-density-wave (CDW) order in kagome metals is highly sensitive to external tuning parameters such as chemical substitution and hydrostatic pressure, which generally suppress long-range order. Here, using high-resolution X-ray diffraction under controlled uniaxial strain, we show that anisotropic lattice deformation instead stabilizes and enhances the CDW state in ScV$_6$Sn$_6$. Compression along the [H00] and [HH0] directions lowers the crystal symmetry from hexagonal to orthorhombic, lifts the degeneracy between symmetry-equivalent in-plane CDW domains, and promotes long-range order while preserving the underlying trimer instability. Phonon calculations indicate only a moderate stabilization of the imaginary flat phonon mode, demonstrating that the increase in T$_\mathrm{CDW}$ is primarily driven by the in-plane ordering of the Sn$^\mathrm{T}$--Sc--Sn$^\mathrm{T}$ \textit{rattling} chains within the frustrated kagome lattice. A phenomenological model incorporating strain-dependent Ising couplings within a three-state Potts framework successfully reproduces the evolution of T$_\mathrm{CDW}$ under compression and captures the continuous nature of the transition. Our results establish uniaxial strain as a powerful symmetry-selective tuning parameter for order-disorder transformations in frustrated lattices.
\end{abstract}

\maketitle
The control and disentanglement of competing electronic orders are central themes in modern condensed matter physics. In quantum materials, such as the cuprate high-$T_{C}$ superconductors or kagome metals, charge-density waves (CDWs), spin order, and superconductivity emerge in close proximity and compete for the same electronic states, giving rise to rich and highly tunable phase diagrams \cite{Keimer_2015,liu2014exotic,hu2022rich,jiang2023kagome,farhang2025discovery,nie2022charge,li2021observation}. In these systems, small perturbations—such as strain, hydrostatic pressure, disorder, or magnetic field—can tip the balance between nearly degenerate ground states \cite{Kim_2018,Gerber_2015,Achkar_2014,Souliou_2018,Yamamoto_2015}, selecting one ordered phase over another. For instance, in layered CDW systems such as 2H-NbSe$_2$, even small uniaxial strain lowers the crystal and electronic symmetry, driving a transition from a triangular (3\textit{Q}) CDW state to a unidirectional (1\textit{Q}) phase \cite{kundu2024charge}.

\begin{figure}
    \centering
    \includegraphics[width=1.0\linewidth]{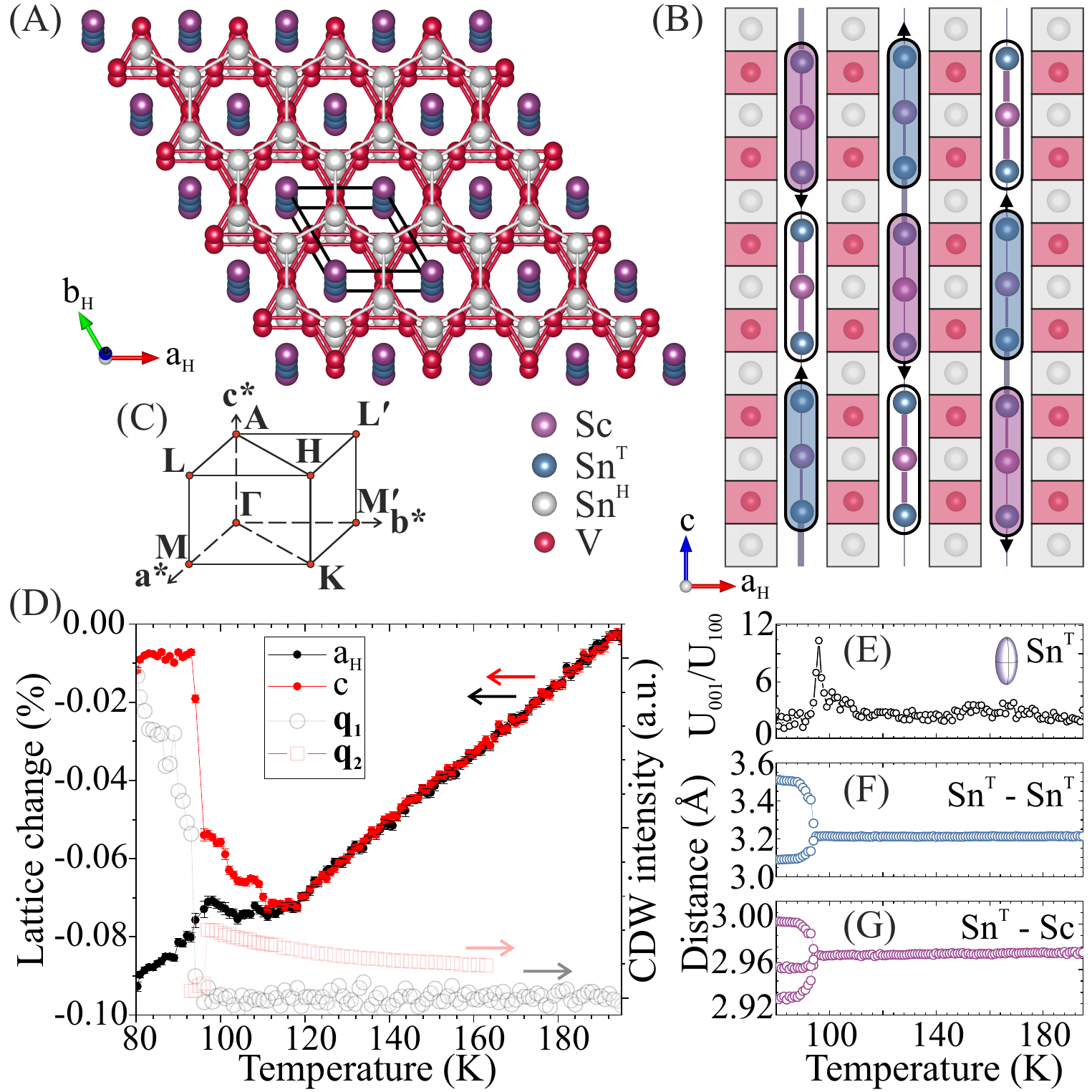}
    \caption{(A) High-temperature crystal structure ($P6/mmm$). V atoms (red) form a kagome lattice, Sn$^\mathrm{H}$ atoms (light grey) occupy honeycomb sites, and Sn$^\mathrm{T}$ (blue) and Sc (purple) form trimers along the out-of-plane direction. (B) CDW-induced structural modulation with the Sn$^\mathrm{T}$--Sc--Sn$^\mathrm{T}$ trimers undergoing collective displacements indicated by arrows. (C) Brillouin zone of the \textit{P6/mmm} structure. (D) Temperature dependence of lattice parameters normalized to their high-temperature values. Translucent symbols indicate the evolution of the integrated intensity of the ${\bf q}_1$ CDW peak (measured in this work) and the ${\bf q}_2$ DS intensity (obtained from \cite{korshunov2023softening}). The intensities are normalized for clarity. (E) Ratio of anisotropic atomic displacement parameters for Sn$^\mathrm{T}$. (F--G) Temperature dependence of the Sn$^\mathrm{T}$--Sn$^\mathrm{T}$ and Sn$^\mathrm{T}$--Sc distances, respectively, indicating site splitting in the modulated phase.}
    \label{Fig1}
\end{figure}

The frustrated nature of the two-dimensional triangular arrangement of transition-metal atoms in the kagome lattice naturally maps the generalized three-state Potts model \cite{enting1982triangular, Adler_1995}, susceptible to order–disorder phase transitions at a critical temperature, $T_\mathrm{C}$ \cite{subires2023order,subires2025frustrated}. This combination of electronic instability and geometric frustration makes kagome metals ideal systems for exploring how competing interactions and degenerate states can be selectively controlled and manipulated, for instance by the application of an in-plane uniaxial strain. By explicitly breaking the rotational symmetry of the kagome lattice, strain acts as a conjugate field to the Ising-like CDW order parameter, energetically favoring one modulation direction over the others. In kagome superconductors of the AV$_3$Sb$_5$ family, uniaxial strain dramatically reshapes the CDW landscape: in CsV$_3$Sb$_5$, for example, compressive strain of $\sim$1\% enhances the CDW gap magnitude by nearly a factor of three \cite{lin2024uniaxial}. Unlike hydrostatic pressure \cite{Stier_2024,Li_2022,Hou_2023}, uniaxial strain introduces directional anisotropy, enabling the selective control of competing ordered states while disentangling intrinsic electronic effects from disorder-driven phenomena.

\begin{figure*}
    \centering
    \includegraphics[width=1.0\linewidth]{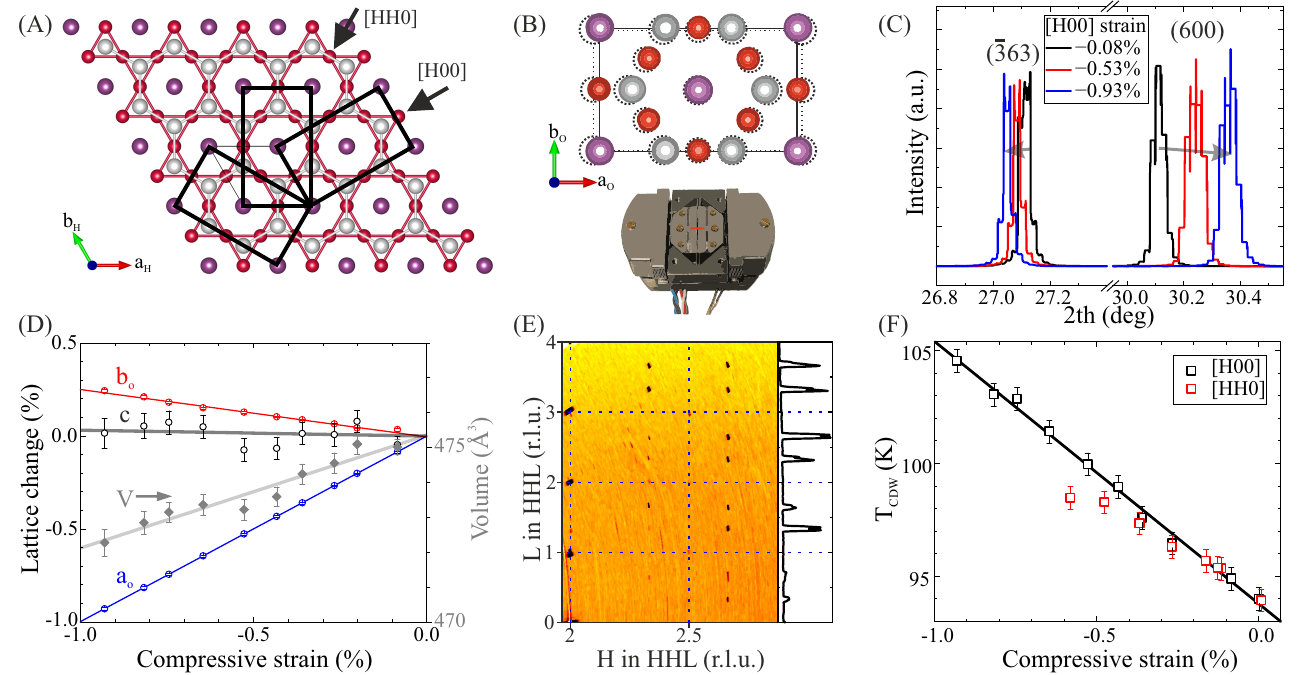}
    \caption{(A) View of the crystal structure along the \textit{c} direction, showing three merohedral orthorhombic domains. Arrows indicate the reciprocal-lattice directions along which uniaxial compressive strain was applied. (B) Upper: sketch of the atomic displacements after the application of uniaxial strain along [H00] direction. Dashed lines indicate the undistorted structure; the final orthorhombic lattice is indicated by solid black lines. Bottom: Sketch of the uniaxial strain device. (C) Shift of characteristic Bragg peaks under applied strain. (D) Evolution of the orthorhombic lattice parameters and volume as a function of applied strain along the [H00] direction measured at $T$ = 107 K. (E) HHL reciprocal-space map measured at $T = 85$ K, showing structural peaks and ${\bf q}_1$ satellites and the intensity distribution along the L direction (side panel). (F) Strain dependence of the CDW transition temperature, $T_\mathrm{CDW}$. Negative strain denotes compression along the indicated parent hexagonal direction.}
    \label{Fig2}
\end{figure*}

The kagome metal ScV$_6$Sn$_6$ provides an ideal platform in which  geometric frustration, lattice symmetry, and electron–phonon coupling compete to generate multiple competing charge-ordered states \cite{Arachchige_2022,Lee_2024,Hu_2025}, rendering the system highly susceptible to uniaxial strain. Despite the pronounced softening of the out-of-plane vibrational mode of the trigonal Sn$^\mathrm{T}$ atoms at ${\bf q}_2=(1/3,\,1/3,\,1/2)$, accompanied by the growth of diffuse scattering intensity upon cooling \cite{korshunov2023softening,Cao_2023}, this instability does not condense into a long-range ordered phase. Rather, the system undergoes a sharp first-order transition into the long-range ${\bf q}_1=(1/3,\,1/3,\,1/3)$ ordered state, driven by the two-dimensional packing of Sn$^\mathrm{T}$--Sc--Sn$^\mathrm{T}$ \textit{rattling} trimers on the triangular kagome lattice \cite{Ganesh_2023}. The strong competition between the nearly degenerate ${\bf q}_1$ and ${\bf q}_2$ instabilities highlights the unconventional nature of the CDW in ScV$_6$Sn$_6$, emphasizing the key role of anharmonicity \cite{wang2024origin} and order–disorder transformation in selecting the ground state. Consequently, the low-temperature $\sqrt{3}~\times~\sqrt{3}~\times~3$ structure can be mapped onto a generalized Ising-like model, in which the positive, neutral, and negative out-of-plane displacements of the trimers define three distinct local states on a triangular lattice \cite{Ganesh_2023, Gomez_2024}, corresponding naturally to a three-state Potts model.

Recent X-ray diffraction studies on ScV$_6$Sn$_6$ have reported a suppression of the CDW under hydrostatic and chemical pressure, and biaxial strain \cite{zhang2022destabilization,meier2023tiny,meier2025pressure,wang2024origin}. This behavior contrasts with the hardening of the soft phonon mode and the enhancement of the CDW transition temperature under uniaxial strain \cite{Tuniz_2025}. Such an increase of T$_\mathrm{CDW}$ is highly unusual among CDW materials \cite{Souliou_2018,Leroux_2015} and suggests a distinct microscopic mechanism underlying the stabilization of the superlattice in ScV$_6$Sn$_6$. Here, we employ X-ray diffraction under uniaxial strain as a symmetry-breaking tuning parameter to disentangle the hierarchy of competing CDW instabilities, and uncover the microscopic origin of the strain-enhanced ordered state. By tracking the structural response under controlled lattice anisotropy, we demonstrate that uniaxial compression lowers the symmetry of the hexagonal lattice toward an orthorhombic structure, lifting the degeneracy between equivalent ordering vectors. In this regime, disorder and trimerization-driven frustration cooperate to stabilize the ${\bf q}_1$ charge-ordered phase. The experimentally observed enhancement of T$_\mathrm{CDW}$ is quantitatively reproduced by Monte Carlo simulations that map the Sn$^\mathrm{T}$--Sc--Sn$^\mathrm{T}$ trimers onto a three-state Potts model with strain-dependent Ising-like couplings \cite{Gomez_2024}. Our results establish uniaxial strain as a powerful probe of emergent Ising-like order in frustrated kagome metals and provide microscopic insight into the mechanisms governing competing charge instabilities in 166 kagome lattices.

We first recall the microscopic nature of the CDW transition in ScV$_6$Sn$_6$ in the absence of external strain, Fig.~\ref{Fig1} and Supplemental Material~\cite{SM}, Sec. II. ScV$_6$Sn$_6$ crystallizes in the hexagonal HfFe$_6$Ge$_6$-type structure, composed of V kagome layers separated by Sn honeycomb layers, Fig.~\ref{Fig1}(A). Below $T_{\mathrm{CDW}}\approx95$~K, the system develops a $\sqrt{3}\times\sqrt{3}\times3$  superstructure \cite{Arachchige_2022} characterized by the ordering vector ${\bf q}_1=(1/3,\,1/3,\,1/3)$. The transition is primarily driven by collective out-of-plane displacements of the Sn$^\mathrm{T}$--Sc--Sn$^\mathrm{T}$ trimers, as illustrated in Fig.~\ref{Fig1}(B). The first-order nature of the transition is evident from the abrupt evolution of both the lattice parameters, and the CDW superlattice peak intensity across $T_\mathrm{CDW}$, Fig.~\ref{Fig1}(D). In addition, the anisotropic displacement parameter ratio, U$_{001}$/U$_{100}$, increases significantly upon cooling due to the large out-of-plane displacements of the Sn$^\mathrm{T}$ atoms, Fig.~\ref{Fig1}(E). Below the transition temperature, the onset of trimerization is reflected in a clear splitting of the Sn$^\mathrm{T}$--Sn$^\mathrm{T}$ bond distances, indicating the freezing of the atomic modulation associated with long-range CDW order, Figs.~\ref{Fig1}(F,G).

To investigate the effect of anisotropic in-plane lattice deformation on the CDW state, we employed in-plane uniaxial strain that explicitly breaks the threefold rotational symmetry of the hexagonal lattice without introducing additional structural disorder, Figs.~\ref{Fig2}(A,B). The strain was applied along the [H00] (or equivalently [HH0]) direction in the hexagonal lattice at T$_{\mathrm{CDW}}$ + 10 K. At this temperature, the lattice is comparatively softer and the competition between the ${\bf q}_1$ and ${\bf q}_2$ instabilities is enhanced. The magnitude of the applied strain was directly quantified from the changes in the refined lattice parameters, manifested as shifts of characteristic Bragg reflections, Fig.~\ref{Fig2}(C). The homogeneous nature of the strain field is evidenced by the absence of Bragg peak splitting or broadening, demonstrating that uniaxial strain provides a clean and controlled route to selectively stabilize a single orthorhombic domain among the three symmetry-equivalent domains of the parent hexagonal lattice, Figs.~\ref{Fig2}(A,B). Our first key observation is the strongly anisotropic lattice response to in-plane compression, as illustrated by the lattice distortion shown in Fig.~\ref{Fig2}(B) along the [H00] direction. More specifically, the orthorhombic $a_o$-axis undergoes substantial compression, while the corresponding $b_o$-axis expands moderately and the \textit{c}-axis remains nearly unchanged, Fig.~\ref{Fig2}(D). The strained phase is well described within the orthorhombic \textit{Cmmm} space group, with lattice parameters satisfying $a_o = \sqrt{3}a_h$, $b_o = b_h$, and $c_o = c_h$ (see details in Supplemental Material~\cite{SM}, Sec. II B). This description is fully consistent with symmetry analysis using ISODISTORT \cite{campbell2006isodisplace, ISODISTORT}, which identifies a single symmetry-allowed orthorhombic distortion associated with the \textit{GM5+} irreducible representation of the parent \textit{P6/mmm} structure. The corresponding CDW structure remains orthorhombic, but the additional superlattice distortion lowers the symmetry relative to the strained high-temperature phase.

\begin{figure*}
    \centering
    \includegraphics[width=1.0\linewidth]{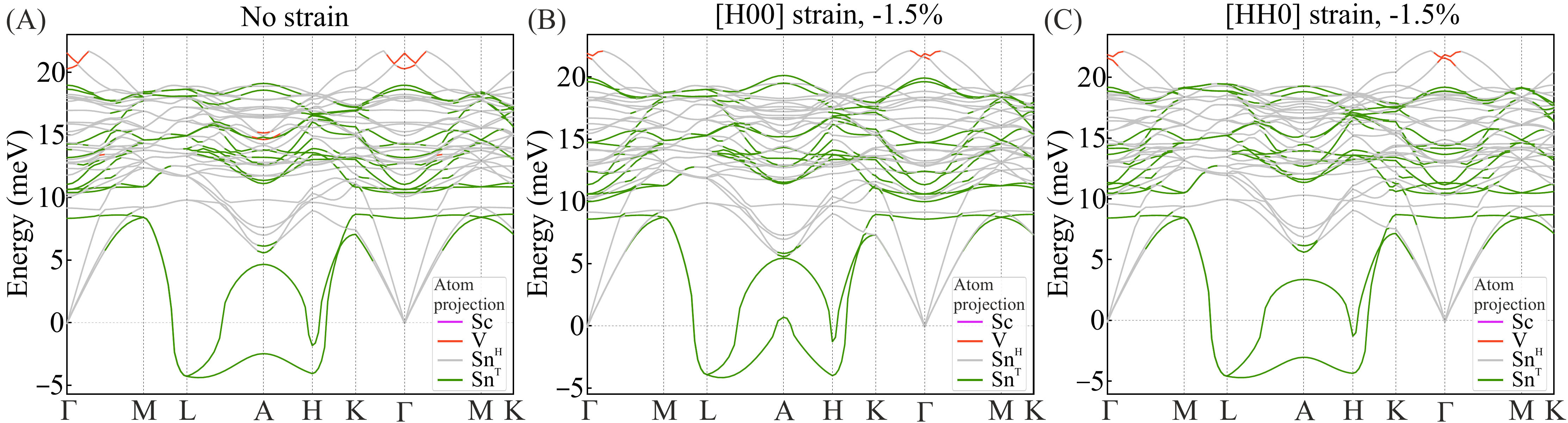}
    \caption{(A--C) Atomic projection of the phonon dispersion of ScV$_6$Sn$_6$ for the (A) no-strain case, (B) 1.5\% compression along [H00], and (C) 1.5\% compression along the [HH0] directions, respectively. The color code describes the different atomic weight on each phonon mode.}
    \label{Fig3}
\end{figure*}

From the temperature dependence of the superlattice reflections, we observe a sharp first-order transition into the charge-ordered state associated with the $\mathbf{q}_1$ modulation, Fig.~\ref{Fig2}(E) and Supplemental Material Sec. III~\cite{SM}. With increasing compressive strain along the [H00] direction, the onset temperature of the CDW intensity shifts monotonically to higher temperatures, Fig.~\ref{Fig2}(F). Uniaxial strain clearly distinguishes between the effects of isotropic volume compression \cite{zhang2022destabilization} and chemical doping \cite{Lee_2024}, manifested here by the enhancement of the CDW phase in ScV$_6$Sn$_6$, despite the overall reduction of the unit-cell volume, Fig.~\ref{Fig2}(D). In addition, we observe no significant enhancement of the competing $\mathbf{q}_2$ modulation. We note that compressive strain applied along the [H00] and [HH0] directions produces qualitatively similar effects, corresponding to compression along the long and short orthorhombic axes, respectively. In both cases, the transition temperature follows a similar strain dependence, indicating only weak anisotropy within the in-plane response, Fig.~\ref{Fig2}(F). This observation imposes constraints on the role of strain-dependent electron–phonon interaction (EPI) effects in stabilizing the $\mathbf{q}_1$ phase or suppressing the competing $\mathbf{q}_2$ instability. In general, one would expect compression along the [HH0] direction to produce a stronger modification of the electron–phonon matrix elements than compression along [H00], leading to distinct strain dependencies of $T_\mathrm{CDW}$. This conclusion is further supported by the calculated strain dependence of the lattice dynamics. Fig.~\ref{Fig3} summarizes the atomic projections of the phonon spectra under compression. The low-energy lattice dynamics is dominated by vibrations of the trigonal (Sn$^\mathrm{T}$) and hexagonal (Sn$^\mathrm{H}$) Sn atoms. In the unstrained lattice, the phonon spectrum exhibits an imaginary nearly flat branch along the L–A–H path  of the BZ \cite{korshunov2023softening,Hu_2025}, corresponding to an out-of-plane vibration of the Sn$^\mathrm{T}$ atoms. The double number of atoms of the orthorhombic unit cell reveals a second Sn$^\mathrm{T}$-derived unstable mode at the L and H points, which becomes stable at the A point, Fig.~\ref{Fig3}(A). Upon compression along either crystallographic direction, the nearly flat negative mode remains unstable at both the L and H points, although it becomes stable at the A point under 1.5\% compression along [H00]. These results indicate that the instability continues to favor the H point as the dominant CDW ordering vector, see Supplemental Material~\cite{SM}, Sec. V.

\begin{figure}
    \centering
    \includegraphics[width=1.0\linewidth]{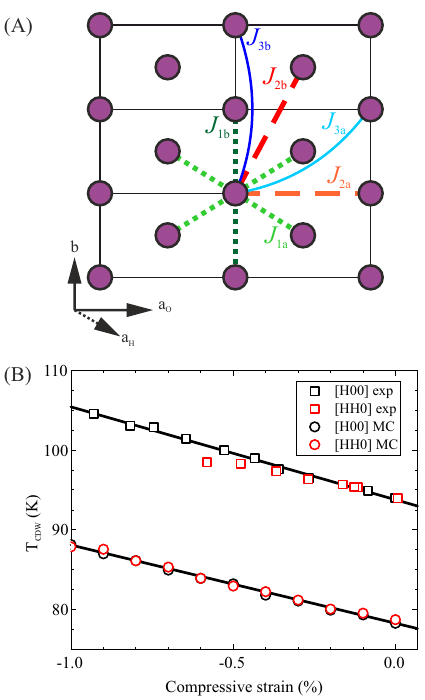}
    \caption{(A) Sketch of the real space definition of the coupling strengths, \textit{J$_i$}, accounting for the first and second nearest neighbors in the orthorhombic cell. \textit{a}$_{h}$ and \textit{b} identify the hexagonal reciprocal lattice vectors. (B) Strain dependence of the transition temperature, $T_\mathrm{CDW}$, extracted from the integrated peak area in the Monte Carlo simulation.}
    \label{Fig4}
\end{figure}

These results raise a fundamental question: is the stability of the CDW order in ScV$_6$Sn$_6$ primarily governed by anharmonic lattice effects or by anisotropic distortions that explicitly break the rotational symmetry of the kagome lattice? Given the absence of the \textit{c}-axis lattice parameter dependence on the strain, it follows that the interaction between the out-of-plane vibration and the \textit{p}$_z$ orbitals of Sn$^\mathrm{T}$ is not affected by the in-plane compression. This, in turn, favors the ordering of the symmetry-equivalent frustrated CDW in the Ising-like triangular kagome lattice \cite{Ganesh_2023,Gomez_2024,yin2022discovery,subires2025frustrated}, generating a manifold of nearly degenerate states. Motivated by these observations, we introduce a phenomenological model in which the interaction strengths (\textit{J$_i$}), illustrated in Fig.~\ref{Fig4}(A) and Supplemental Material~\cite{SM}, Sec.VI, explicitly depend on the applied strain according to: 

\begin{equation}
    J_i(d) = J_{i0} \times \exp\left[\lambda \left(1 - \frac{d}{d_0}\right)\right]
\end{equation}  

where $J_{i0}$ is the interaction parameter for the unstrained system and $d_0$ is the interatomic distance in the unstrained system and $\lambda$ accounts for a scaling factor ($\lambda = 25$). In the orthorhombic phase, where compression is applied along the \textit{a}$_{o}$=$<$210$>$$_{h}$ direction, the strain dependence of the coupling parameters \textit{J} is encoded through the evolution of the corresponding bond lengths \textit{l}$_{j´s}$ (see details in Supplemental Material~\cite{SM}, Sec. VI). Minimization of the Hamiltonian and further Fourier transform of the real space trimer distribution reproduces the $\sqrt{3}\times\sqrt{3}$ diffuse scattering, Fig.~S11 \cite{neder2008diffuse}. The temperature dependence reveals an increase of the transition temperature within the experimentally accessible strain range, Fig.~\ref{Fig4}(B), demonstrating that the transition temperature is highly sensitive to the strength of the in-plane coupling parameters of the effective Ising Hamiltonian. More importantly, the simulations support a scenario in which the enhancement of the CDW under compressive strain originates from the stabilization and ordering of the Sn$^\mathrm{T}$--Sc--Sn$^\mathrm{T}$ trimers through an order-from-disorder mechanism, rather than from Fermi-surface nesting effects or changes in the electron–phonon scattering matrix elements. Within this framework, the competing nature of the charge modulations naturally leads to the suppression of the $\mathbf{q}_2$ instability, as strain selectively favors and pins local $\mathbf{q}_1$ CDW domains, thereby hardening the associated soft phonon mode \cite{Tuniz_2025}. This interpretation is further supported by the evolution of the phase transition itself. In the absence of strain, the $\mathbf{q}_1$ modulation exhibits a pronounced first-order transition. Under compression, however, trimer disorder progressively smears the sharp crossover, and the order parameter evolves toward a more continuous transition, as expected for a three-state Potts model (Supplemental Material~\cite{SM}, Fig.~S3).

We would like to briefly mention the consequences of the hexagonal to orthorhombic distortion in kagome metals. As a very small amount of strain is sufficient to break the \textit{P6/mmm} symmetry, the rotational symmetry is reduced from \textit{C}$_6$ to \textit{C}$_2$. Besides lifting the degeneracy between the van Hove singularity and Dirac points at M (one nesting channel becomes dominant) and K, the 3 equivalent directions split and favor a single-\textbf{q} CDW wavevector. This, in turn, is detrimental for chiral phases, but favors nematic  --strain acts as a conjugate field to nematicity-- and stripe order \cite{farhang2025discovery,Jiang_2024}. This brings important consequences for (Cs,K,Rb)V$_3$Sb$_5$, where the ground state is being debated between chiral and nematic orders \cite{Guo_2022,Guo_2024,Guo_2024a,Elmers_2025,Jiang_2021,Mielke_2022,sadrollahi_2025,Jiang_2023} or whether a chiral-to-nematic transition is present provided that the amplitudes and phases evolve continuously with strain.
  
In conclusion, we demonstrated that the enhancement of the CDW under uniaxial strain in the non-magnetic kagome metal ScV$_6$Sn$_6$ originates from the long-range ordering of the $c$-axis–oriented Sn$^\mathrm{T}$--Sc--Sn$^\mathrm{T}$ \textit{rattling} chains, rather than from purely electronic- or lattice-driven mechanisms. The strain dependence of $T_\mathrm{CDW}$ is well captured by Monte Carlo simulations that incorporate anisotropic, strain-dependent Ising coupling strengths within a three-state Potts model. These results provide a pathway to disentangle competing and coexisting order parameters in multi-q CDW kagome metals and establish a new framework for controlling order–disorder transitions in frustrated triangular lattices.

\section*{Acknowledgments}
We thank Sajna Hameed, Alexander Sukhanov, Johannes Frey, and ESRF sample environment group for valuable discussions and their assistance with the preparation of the experimental setup. A.K. thanks the Basque government for financial support through the project PIBA-2023-1-0051, and S.B-C. to the MINECO of Spain through the projects PID2021-122609NB-C21 and PID2024-161503NB-C21.  V.P. acknowledges support from the Ministry of Science of Spain through the Projects No. PID2021-122609NB-C22 and PID2024-161503NB-C22. C.-Y.L. is supported by the European Research Council (ERC) under the European Union’s Horizon 2020 research and innovation program (grant agreement no. 101020833) and by the European Union Next Generation EU/PRTR-C17.I1, as well as by IKUR Strategy under the collaboration agreement between Ikerbasque Foundation and DIPC on behalf of the Department of Education of the Basque Government.  VP and JCS acknowledge support from the Ministry of Science of Spain through the Projects No. PID2021-122609NB-C22 and PID2024-161503NB-C22. We thank the CESGA (Centro de Supercomputacion de Galicia) for the computing facilities provided. J.C.S. thanks the support of the Ministry of Science and Education through the FPU Program (FPU22/01312). C.Y., C.S., and C.F. acknowledge financial support by the Deutsche Forschungsgemeinschaft  (DFG, German Research Foundation) through the SFB 1143 (project ID 247310070), QUAST (project ID FOR 5249), the  Würzburg-Dresden Cluster of Excellence ctd.qmat – Complexity, Topology and Dynamics in Quantum Matter (EXC 2147, project-id 390858490), and the European Union through EXQIRAL (No. 101131579). This research used the ARPES end station of the ESM beamline of the National Synchrotron Light Source II, a U.S. Department of Energy (DOE) Office of Science User Facility operated for the DOE Office of Science by Brookhaven National Laboratory under Contract No. DE-SC0012704.

\bibliography{biblio}

\end{document}


\title{Supplemental Material for: Symmetry-Selective Stabilization of Charge-Density Wave in ScV$_6$Sn$_6$}

\author{A. Korshunov}
\thanks{These authors contributed equally to this work.}
\affiliation{Donostia International Physics Center (DIPC), Paseo Manuel de Lardizábal. 20018, San Sebastián, Spain}

\author{C.-Y. Lim}
\thanks{These authors contributed equally to this work.}
\affiliation{Donostia International Physics Center (DIPC), Paseo Manuel de Lardizábal. 20018, San Sebastián, Spain}

\author{J. Corral-Sertal}
\thanks{These authors contributed equally to this work.}
\affiliation{Departamento de Física Aplicada, Universidade de Santiago de Compostela, E-15782 Campus Sur s/n, Santiago de Compostela, Spain}
\affiliation{CiQUS, Centro Singular de Investigacion en Quimica Biolóxica e Materiais Moleculares, Departamento de Quimica-Fisica, Universidade de Santiago de Compostela, Santiago de Compostela, E-15782, Spain}

\author{G. Garbarino}
\affiliation{European Synchrotron Radiation Facility (ESRF), BP 220, F-38043 Grenoble Cedex, France}

\author{D. Chernyshov}
\affiliation{Swiss-Norwegian BeamLines at European Synchrotron Radiation Facility, BP 220, F-38043 Grenoble Cedex, France}

\author{A. Rajapitamahuni}
\affiliation{National Synchrotron Light Source II, Brookhaven National Laboratory, Upton, New York 11973, USA}
\affiliation{Department of Physics, SRM University - AP, Amaravati, Andhra Pradesh, 522502,India.}
\affiliation{Department of Physics and Astronomy, University of Nebraska-Lincoln, Nebraska 68588, USA.}

\author{C. Yi}
\affiliation{Max Planck Institute for Chemical Physics of Solids, D-01187 Dresden, Germany}

\author{ S. Roychowdhury}
\affiliation{Max Planck Institute for Chemical Physics of Solids, D-01187 Dresden, Germany}
\affiliation{Department of Chemistry, Indian Institute of Science Education and Research Bhopal, Bhopal-462 066, India}

\author{C. Shekhar}
\affiliation{Max Planck Institute for Chemical Physics of Solids, D-01187 Dresden, Germany}

\author{C. Felser}
\affiliation{Max Planck Institute for Chemical Physics of Solids, D-01187 Dresden, Germany}

\author{V. Pardo}
\affiliation{Departamento de Física Aplicada, Universidade de Santiago de Compostela, E-15782 Campus Sur s/n, Santiago de Compostela, Spain}
\affiliation{Instituto de Materiais iMATUS, Universidade de Santiago de Compostela, E-15782 Campus Sur s/n, Santiago de Compostela, Spain} 

\author{Ella M. Schmidt}
\affiliation{MARUM - Center for Marine Environmental Sciences, University of Bremen, 28359, Bremen, Germany}
\affiliation{Crystallography and Geomaterial Research, Faculty of Geosciences, University of Bremen, 28359, Bremen, Germany}
\affiliation{MAPEX Center for Materials and Processes, University of Bremen, 28359, Bremen, Germany}

\author{S. Blanco-Canosa}
\email{sblanco@dipc.org}
\affiliation{Donostia International Physics Center (DIPC), Paseo Manuel de Lardizábal. 20018, San Sebastián, Spain}
\affiliation{IKERBASQUE, Basque Foundation for Science, 48013 Bilbao, Spain}

\date{\today}

\maketitle

\section{Experimental details}
\subsection{Sample characterization}

High-quality single crystals of ScV$_6$Sn$_6$ were grown by a flux method, as described in \cite{korshunov2023softening}. X-ray single-crystal diffraction measurements were carried out at the BM01 beamline (Swiss--Norwegian Beamlines, SNBL) of the European Synchrotron Radiation Facility (ESRF) using an incident energy of 17.5~keV (0.71 \r{A}) and a Dectris PILATUS 2M area detector~\cite{dyadkin2016new}. Low-temperature measurements spanning the CDW transition (80--200~K) were performed using a Cryostream N$_2$ blower (Oxford Cryosystems). Diffraction images were collected with a 1$^\circ$ step in angular rotation over a full 360$^\circ$ range. The raw data were averaged over five repeated scans to improve statistics and further processed using the \textit{SNBL Toolbox}. Sequential unit cell indexing and Bragg peaks integration at different temperatures were carried out in \textit{CrysAlisPro} software and structural refinement was performed using \textit{SHELX} \cite{sheldrick2008short}.

\subsection{Uniaxial strain}

Single-crystal diffraction measurements under uniaxial strain were performed at the ID15B beamline (ESRF) \cite{garbarino2024extreme}. A bulk single crystal of ScV$_6$Sn$_6$ was saw-cut into plates with the $c$ axis perpendicular to the surface and subsequently polished to a thickness of approximately 100~\textmu m. Thin, bar-shaped samples with a cross section of approximately 2~mm $\times$ 150~\textmu m were cut using laser drilling, such that uniaxial strain could be applied along the [H00] (or [HH0]) direction oriented along the long axis of the bar. The samples were subsequently chemically etched to a final thickness of approximately 70~\textmu m and mounted on a Razorbill CS200T strain device using Loctite Stycast 2850FT epoxy with CAT 24LV catalyst. The strain device, with the sample mounted horizontally, was placed in a cryostat to allow full access to Bragg reflections along the strain direction. An incident energy of $E_\mathrm{i} = 30$~keV ($\lambda = 0.41$~\AA) was used, and the scattered photons were collected using an EIGER2 X 9M CdTe detector. Diffraction images were recorded over an angular range of $\pm 35^\circ$ with an angular step of 0.5$^\circ$. Raw data were processed using \textit{CrysAlisPro}, and structural refinement was performed in \textit{Jana2006} \cite{Jana2006}.

\subsection{Angle-Resolved Photoemission Spectroscopy (ARPES)}

ARPES experiments were performed at the 21-ID ESM beamline at NSLS-II using a Scienta DA30 analyzer. The samples were \textit{in-situ} cleaved inside an ultra-high-vacuum chamber with a base pressure better than ~$4 \times10^{-11}$ Torr and maintained at the measurement temperature. A photon energy of 107 eV was used, corresponding to the $\Gamma$ plane ($k_z$ = 0), and the polarization conditions are specified in the figure captions. The energy and momentum resolutions were better than 10 meV and 0.01 \r{A}$^{-1}$, respectively. 

\section{Crystal structure}
\subsection{High-temperature phase}

\begin{figure}[t]
    \centering
    \includegraphics[width=\linewidth]{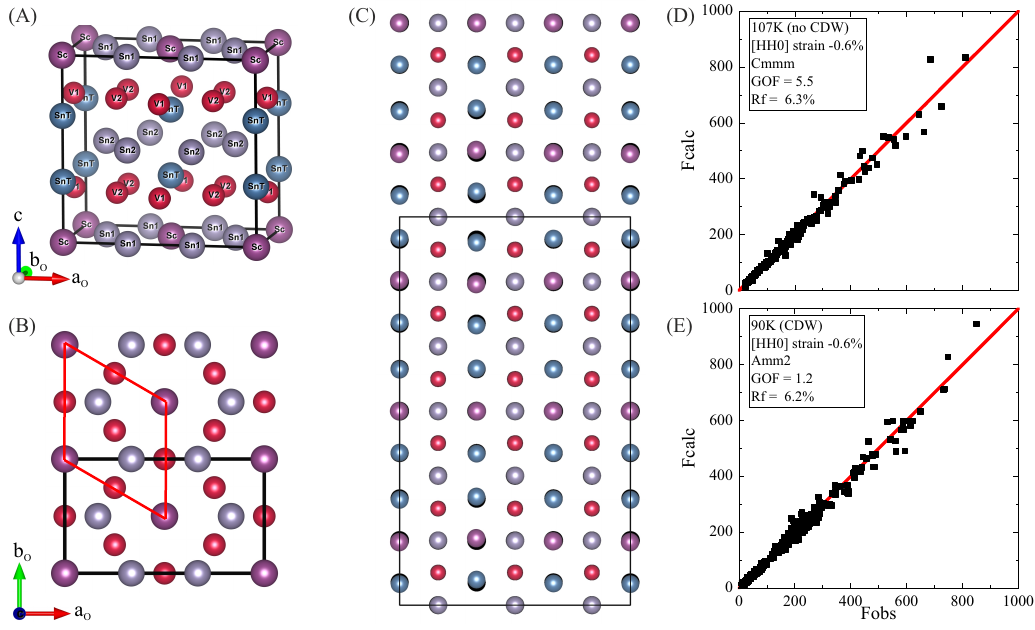}
    \caption{Strain-induced orthorhombic structure in the \(Cmmm\) setting above the phase transition, shown in a common view (A) and along the $c$ direction (B). The parent hexagonal unit cell is outlined in red in (B). 
    (C) Refined CDW modulation in the \(Amm2\) space group. Displaced atoms are colored, while the nonmodulated atomic positions are shown as black shadows. 
    (D,E) Comparison between the calculated structure factors \(F_{\mathrm{calc}}\) and the experimental values \(F_{\mathrm{obs}}\) for the observed Bragg reflections, demonstrating the quality of the refinement for the high-temperature \(Cmmm\) phase and the low-temperature \(Amm2\) phase, respectively.
    }
    \label{fig:S1_structure}
\end{figure}

The high-temperature phase of \(\mathrm{ScV_6Sn_6}\) crystallizes in the hexagonal \(P6/mmm\) space group (191)~\cite{korshunov2023softening}. No splitting of the main Bragg reflections indicative of an orthorhombic distortion was observed in the present work or reported previously. The crystal structure consists of ordered layered fragments formed by: (i) trigonal Sc layers and hexagonal \(\mathrm{Sn}^{H}\), (ii) kagome V and trigonal \(\mathrm{Sn}^{T}\), and (iii)  hexagonal \(\mathrm{Sn}^{H}\) layers, Fig.~S\ref{fig:S1_structure}A. The V and \(\mathrm{Sn}^{H}\) atoms form a relatively rigid hollow framework, while the central void is occupied by the quasi-one-dimensional \(\cdots\mathrm{Sn}^{T}\!-\!\mathrm{Sc}\!-\!\mathrm{Sn}^{T}\cdots\) chain running along the \(c\) axis.

\subsection{Strain-induced orthorhombic distortion}

\begin{figure}[t]
    \centering
    \includegraphics[width=\linewidth]{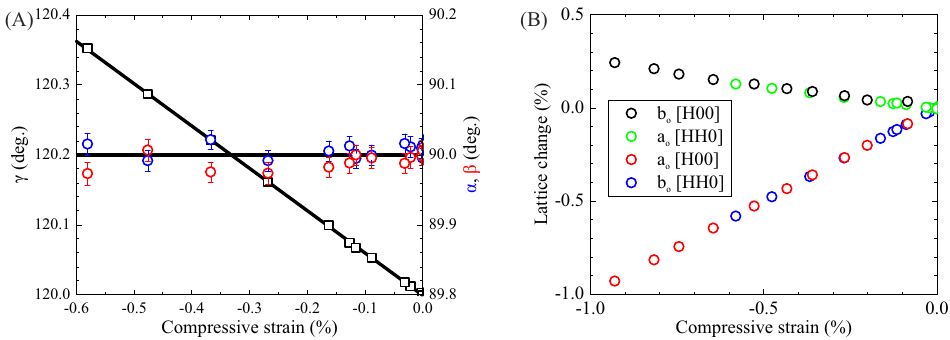}
    \caption{
    Strain-induced evolution of \(\mathrm{ScV_6Sn_6}\) lattice.
    (A) Evolution of the unit cell angles in pseudo-hexagonal setting under applied in-plane uniaxial strain. 
    (B) Strain dependence of the orthorhombic lattice parameters for compression applied along the parent hexagonal [H00] and [HH0] directions.
    }
    \label{fig:SI_strain}
\end{figure}

Application of in-plane uniaxial strain induces a homogeneous structural distortion of the high-temperature phase, resulting in characteristic shifts of the main Bragg reflections. In particular, while the \(\alpha\) and \(\beta\) angles in the pseudo-hexagonal setting remain close to \(90^\circ\), the \(\gamma\) angle deviates significantly from \(120^{\circ}\), and the structure is continuously transformed into an orthorhombic setting, as shown in Fig.~S\ref{fig:SI_strain}.

A symmetry-mode analysis performed using ISODISTORT shows that the strain-induced distortion is described by a single $GM5+$ irreducible representation associated with the symmetry lowering from hexagonal \(P6/mmm\) to orthorhombic \(Cmmm\) space group. The applied uniaxial strain selects a single merohedral orthorhombic domain, Fig.~S\ref{fig:S1_structure}B, whose lattice is related to the parent hexagonal basis by
\[
\begin{pmatrix}
\mathbf{a}_{o} \\
\mathbf{b}_{o} \\
\mathbf{c}_{o}
\end{pmatrix}
=
\begin{pmatrix}
1 & 0 & 0 \\
1 & 2 & 0 \\
0 & 0 & 1
\end{pmatrix}
\begin{pmatrix}
\mathbf{a}_{h} \\
\mathbf{b}_{h} \\
\mathbf{c}_{h}
\end{pmatrix}.
\]
where the subscripts \(o\) and \(h\) denote the orthorhombic and hexagonal settings, respectively.

The orthorhombic structure preserves the same chemical composition as the parent phase, with one Sc atom, two \(\mathrm{Sn}^{H}\) atoms, and one \(\mathrm{Sn}^{T}\) atom per formula unit. Although the reduced symmetry allows a splitting of the kagome V position along the $c$-direction, the corresponding displacement is negligible within the experimental uncertainty. The refined atomic coordinates and displacement parameters are summarized in Table~\ref{tab:strained_highT_structure}.

\begin{table*}[t]
\caption{
Refined structural parameters of \(\mathrm{ScV_6Sn_6}\) measured under uniaxial [HH0] compressive strain of \(-0.6\%\) at \(107~\mathrm{K}\). The structure was refined in the orthorhombic \(Cmmm\) space group, with lattice parameters \(a_{o}=9.4751(2)~\text{\AA}\), \(b_{o}=5.4311(2)~\text{\AA}\), \(c_{o}=9.155(5)~\text{\AA}\), and unit-cell volume \(V=471.2(4)~\text{\AA}^{3}\). The refinement residuals are \(R_{F}=6.3\%\), \(wR_{F}=8.1\%\), and \(\mathrm{GOF}=5.5\).
}
\label{tab:strained_highT_structure}
\centering
\begin{tabular}{lccccc}
\toprule
Atom & Occ. & \(x\) & \(y\) & \(z\) & \(U_{\mathrm{iso}}\)~(\(\text{\AA}^{2}\)) \\
\midrule
Sn$^T$ & 1 & 0   & 0           & 0.6759(7)  & 0.011(4)    \\
Sn1 & 1 & 0.5 & 0.1664(2) & 0          & 0.013(3)    \\
Sn2 & 1 & 0.5 & 0.1663(2) & 0.5        & 0.011(3)    \\
V1  & 1 & 0.5 & 0           & 0.7517(14) & 0.010(9)   \\
V2  & 1 & 0.75& 0.25        & 0.2483(9)  & 0.009(1)   \\
Sc & 1 & 0   & 0           & 0          & 0.012(1)  \\
\bottomrule
\end{tabular}
\end{table*}

\subsection{Uniaxial-strain dependence of the lattice parameters}
The displacement applied by the strain device is distributed among the sample carrier, the epoxy, and the sample itself. Consequently, without an independent calibration, the strain experienced by the sample cannot be determined directly from either the voltage applied to the piezo stacks or the displacement measured by the capacitive sensor alone. X-ray diffraction provides a distinct advantage in this respect, as it directly probes the crystal lattice through the positions of the Bragg reflections, allowing the lattice parameters, and therefore the strain state of the sample, to be determined in situ.
In the present experiment, all diffraction peaks were indexed in the orthorhombic setting, and the lattice parameters were refined accordingly. Since the sample remained continuously mounted in the strain device during cooling, a finite residual strain could arise from differential thermal contraction and the residual stroke of the piezo stacks. Assuming that the in-plane lattice parameters are equal in the unstrained state, the zero-strain condition at low temperature was estimated by a linear extrapolation of the refined lattice parameter $a_h$ as a function of the orthorhombic distortion $a_o/\sqrt{3} - b_o$. We find that the zero-strain value estimated in this manner agrees closely with the value measured directly after mechanically breaking the sample, leaving the crystal attached to only one side of the device and therefore effectively strain free.

As noted above, applying uniaxial strain along a selected in-plane direction allows a single merohedral orthorhombic domain to be preferentially populated. We find that the lattice response is equivalent for [HH0] and [H00] uniaxial strain: the lattice parameters evolve in the same way within experimental uncertainty, as shown in Fig.~S\ref{fig:SI_strain}B. However, compared to the orthorhombic unit cell (Fig. 2B of the main text), the direction is opposite: for [H00] strain, compression is applied along the long orthorhombic axis, whereas [HH0] strain corresponds to compression along the short orthorhombic axis.

\subsection{Low-temperature CDW phase}

At ambient pressure, the CDW phase of \(\mathrm{ScV_6Sn_6}\) is characterized by a \(\sqrt{3}\times\sqrt{3}\times3\) superstructure induced by propagation vector ${\bf q}_1=(1/3,\,1/3,\,1/3)$
in the parent hexagonal basis. The primary structural distortion is associated with a trimerization of the \(\mathrm{Sn}^{T}\!-\!\mathrm{Sc}\!-\!\mathrm{Sn}^{T}\) chains, in which consecutive trimers shift along the \(c\) direction to form the ordered pattern shown in Fig.~1 of the main text. Under uniaxial strain, the CDW transition appears to originate from the same structural instability, see Fig.~S\ref{fig:S1_structure}C. The diffraction pattern remains qualitatively similar to that observed at ambient pressure, and the satellite reflections can be indexed by the propagation vector ${\bf q}_o=(0,\,1/3,\,1/3)$ in the orthorhombic basis.

Group--subgroup analysis allows several orthorhombic subgroups of the parent \(P6/mmm\) structure, including \(Cmmm\), \(C222\), \(Cmm2\), and \(Amm2\). A comparison of the corresponding refinements is given in Table~\ref{tab:refinement_model_comparison}. Among these possibilities, the most stable refinement was obtained in the \(Amm2\) space group. Although this model requires a larger number of independent atomic sites and refined parameters (Table~\ref{tab:strained_CDW_structure}) than the strained high-temperature \(Cmmm\) structure, it provides a reasonably good fit. The refinement remains stable even when anisotropic displacement parameters are used for all atoms, which further improve refined R-factors. The refinement clearly resolves sizeable displacements of trimers, formed by \(\mathrm{Sn}^{T}\) and Sc atoms, as shown in Fig.~S\ref{fig:S1_structure}C. This supports a close structural similarity between the strain-induced CDW phase and the CDW structure observed at ambient pressure.

\begin{table*}[t]
\centering
\caption{
Comparison of structural refinement results for different symmetry models of the strained CDW phase of \(\mathrm{ScV_6Sn_6}\).
}
\label{tab:refinement_model_comparison}
\begin{tabular}{lccccc}
\toprule
 & \(Cmmm\) & \(Amm2\) & \(Amm2\) (aniso) & \(Cmm2\) & \(C222\) \\
\midrule
SG number & 65 & 38 & 38 & 35 & 21 \\
GOF & 5.20 & 1.19 & 1.06 & 1.76 & 2.64 \\
\(R_{F}\), \% & 17.05 & 6.22 & 5.17 & 12.10 & 20.45 \\
\(wR_{F}\), \% & 33.74 & 9.67 & 8.03 & 8.19 & 8.41 \\
Number of atoms & 30 & 45 & 45 & 54 & 38 \\
Number of parameters & 81 & 112 & 310 & 103 & 118 \\
Observed reflections, \(I > 3\sigma(I)\) & 869 & 1405 & 1405 & 994 & 1266 \\
\bottomrule
\end{tabular}
\end{table*}

\begin{table*}[t]
\centering
\caption{
Refined structural parameters of $\mathrm{ScV_6Sn_6}$ measured under uniaxial [HH0] compressive strain of $-0.6\%$ at $90~\mathrm{K}$. The structure was refined in the orthorhombic $Amm2$ space group (38), with lattice parameters $a_{o}=27.514(19)~\text{\AA}$, $b_{o}=9.4667(7)~\text{\AA}$, $c_{o}=16.2832(9)~\text{\AA}$, and unit-cell volume $V=4241(3)~\text{\AA}^{3}$. Isotropic displacement parameters were used for all atoms. The refinement residuals are $R_{F}=6.22\%$, $wR_{F}=9.67\%$, and $\mathrm{GOF}=1.19$.
}
\label{tab:strained_CDW_structure}
\begin{tabular}{lccccc}
\toprule
Atom & Occ. & \(x\) & \(y\) & \(z\) & \(U_{\mathrm{iso}}\)~(\(\text{\AA}^{2}\)) \\
\midrule
Sn1-1 & 1 & 0.7285(3) & 0.0000 & 0.0000(1) & 0.0055(2) \\
Sn1-2 & 1 & 0.7201(2) & 0.0000 & 0.3324(1) & 0.0057(2) \\
Sn1-3 & 1 & 0.7269(3) & 0.0000 & 0.6668(1) & 0.0059(3) \\
Sn1-4 & 1 & 0.0540(3) & 0.0000 & -0.0002(1) & 0.0050(2) \\
Sn1-5 & 1 & 0.0659(1) & 0.0000 & 0.3328(1) & 0.0058(2) \\
Sn1-6 & 1 & 0.0560(4) & 0.0000 & 0.6667(1) & 0.0056(3) \\
Sn1-7 & 1 & 0.3931(2) & 0.0000 & -0.0004(1) & 0.0073(3) \\
Sn1-8 & 1 & 0.3902(1) & 0.0000 & 0.3332(1) & 0.0061(2) \\
Sn1-9 & 1 & 0.3924(2) & 0.0000 & 0.6666(1) & 0.0055(3) \\
Sn2-1 & 1 & 0.8334(1) & 0.3344(1) & 0.3335(1) & 0.0063(3) \\
Sn2-2 & 1 & 0.8328(1) & 0.8337(1) & 0.1666(1) & 0.0068(3) \\
Sn2-3 & 1 & 0.8338(1) & 0.3336(1) & 0.0005(1) & 0.0063(3) \\
Sn2-4 & 1 & 0.5000 & 0.8330(1) & 0.8328(2) & 0.0071(3) \\
Sn2-5 & 1 & 0.5000 & 0.3330(1) & 0.6665(1) & 0.0082(3) \\
Sn2-6 & 1 & 0.5000 & 0.3336(1) & -0.0005(1) & 0.0066(3) \\
Sn3-1 & 1 & 0.0000 & 0.8325(1) & 0.8334(2) & 0.0068(3) \\
Sn3-2 & 1 & 0.0000 & 0.3346(1) & 0.6681(1) & 0.0057(3) \\
Sn3-3 & 1 & 0.0000 & 0.3345(2) & -0.0004(1) & 0.0086(3) \\
Sn3-4 & 1 & 0.3335(1) & 0.3348(1) & 0.3339(1) & 0.0065(3) \\
Sn3-5 & 1 & 0.3329(1) & 0.8332(1) & 0.1666(1) & 0.0077(3) \\
Sn3-6 & 1 & 0.3337(1) & 0.3335(1) & 0.0007(1) & 0.0068(3) \\
\midrule
V1  & 1 & 0.7508(2) & 0.7500(3) & 0.0831(3) & 0.0058(4) \\
V2  & 1 & 0.7505(2) & 0.7508(3) & 0.4161(3) & 0.0066(4) \\
V3  & 1 & 0.7506(2) & 0.2493(3) & 0.2497(3) & 0.0060(4) \\
V4  & 1 & 0.0840(2) & 0.7500(3) & 0.0828(3) & 0.0062(4) \\
V5  & 1 & 0.0837(2) & 0.7496(3) & 0.4164(3) & 0.0064(4) \\
V6  & 1 & 0.0842(2) & 0.2503(3) & 0.2496(3) & 0.0065(4) \\
V7  & 1 & 0.4176(2) & 0.7498(3) & 0.0829(3) & 0.0062(4) \\
V8  & 1 & 0.4176(2) & 0.7497(3) & 0.4163(3) & 0.0063(4) \\
V9  & 1 & 0.4172(2) & 0.2501(3) & 0.2496(3) & 0.0063(4) \\
V10 & 1 & 0.7510(3) & 0.0000 & 0.8347(4) & 0.0084(7) \\
V11 & 1 & 0.7508(3) & 0.0000 & 0.1686(4) & 0.0078(6) \\
V12 & 1 & 0.7505(4) & 0.0000 & 0.5012(4) & 0.0086(7) \\
V13 & 1 & 0.0838(3) & 0.0000 & 0.8349(4) & 0.0085(6) \\
V14 & 1 & 0.0843(3) & 0.0000 & 0.1679(4) & 0.0082(7) \\
V15 & 1 & 0.0838(3) & 0.0000 & 0.5015(4) & 0.0084(7) \\
V16 & 1 & 0.4182(3) & 0.0000 & 0.8348(4) & 0.0081(6) \\
V17 & 1 & 0.4172(3) & 0.0000 & 0.1680(4) & 0.0087(7) \\
V18 & 1 & 0.4182(3) & 0.0000 & 0.5016(4) & 0.0082(7) \\
\midrule
Sc1 & 1 & 0.8371(4) & 0.0000 & -0.0020(3) & 0.0070(6) \\
Sc2 & 1 & 0.8269(2) & 0.0000 & 0.3314(3) & 0.0052(6) \\
Sc3 & 1 & 0.8352(4) & 0.0000 & 0.6647(3) & 0.0067(7) \\
Sc4 & 1 & 0.5000 & 0.0000 & -0.0022(4) & 0.0114(9) \\
Sc5 & 1 & 0.5000 & 0.0000 & 0.3341(3) & 0.0071(7) \\
Sc6 & 1 & 0.5000 & 0.0000 & 0.6647(4) & 0.0125(9) \\
\bottomrule
\end{tabular}
\end{table*}

\section{CDW transition temperature}

\begin{figure}
	\centering
	\includegraphics[width=\linewidth]{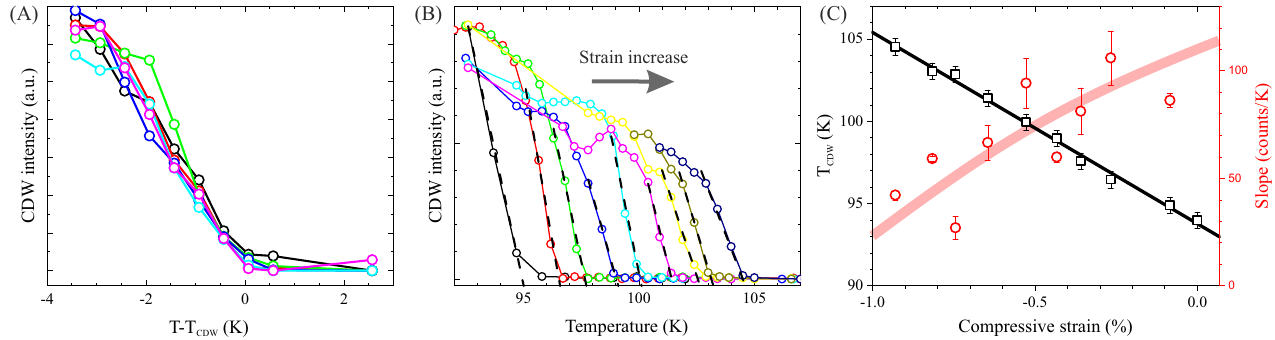}
	\caption{
    (A) Temperature dependence of several CDW peaks located along and perpendicular to the strain direction in the vicinity of the phase transition.
    (B) Temperature dependence of the CDW intensity for the same peak at different values of applied [H00] compressive strain. The dashed lines correspond to linear fits to the intensity near the transition.
    (C) [H00] compressive strain dependence of the CDW transition temperature and of the fitted slope.
    }
	\label{fig:slope}
\end{figure}

To track the CDW intensity under applied strain, high-intensity datasets were collected at ID15B by adjusting the undulator gap. For each dataset, all rotation frames were summed and subsequently analyzed. We find that the CDW reflections located both along and perpendicular to the strain direction exhibit nearly the same temperature dependence for all applied strain values, Fig.~S\ref{fig:slope}A.
The CDW peak intensities were obtained by integrating the scattering intensity within a small region around each satellite reflection and were plotted as a function of temperature for different strain values, Fig.~S\ref{fig:slope}B. Since the CDW transition is relatively sharp, the transition temperature was estimated by fitting the integrated CDW intensity with a linear function in the vicinity of the phase transition. With increasing strain, we observe an overall reduction of the saturation intensity, accompanied by a gradual broadening of the transition, which is reflected in a small change of the fitted slope, Fig.~S\ref{fig:slope}C. 

\section{ARPES}

\begin{figure}
	\centering
	\includegraphics{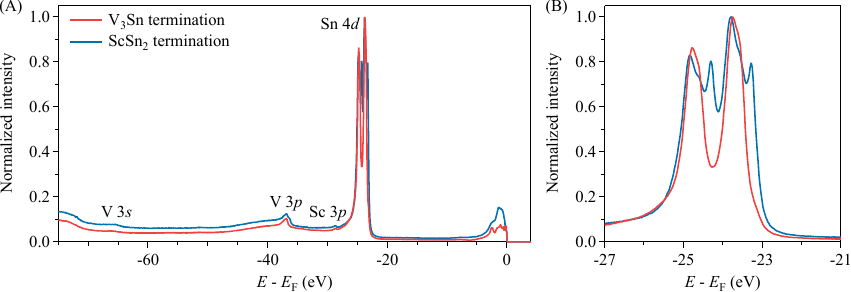}
	\caption{
    Core-level spectra of ScV$_6$Sn$_6$ measured with 107 eV photons.
    (A) Full spectra with Sc 3$p$, V 3$s$, V 3$p$, and Sn 4$d$ core-level peaks. Red and blue curves correspond to the V$_3$Sn and ScSn$_2$ terminations, respectively. 
    (B) Zoomed-in spectra of (A) near the Sn 4$d$ peaks.
    }
	\label{fig:ARPES1}
\end{figure}

Prior to the electronic structure investigation with ARPES, core-level spectra of two different surface terminations were measured using 107 eV photons, as shown in Fig.~S\ref{fig:ARPES1}. The full spectra, Fig.~S\ref{fig:ARPES1}A, exhibit clear Sc 3$p$, V 3$s$, V 3$p$, and Sn 4$d$ core-level peaks, confirming the elemental composition of the samples. The Sn 4$d$ peaks in Fig.~S\ref{fig:ARPES1}B clearly distinguish the V$_3$Sn-terminated (red) and ScSn$_2$-terminated (blue) surfaces. The two broad peaks at approximately -23.5 eV and -24.5 eV correspond to the spin-orbit-split Sn 4$d_{5/2}$ and 4$d_{3/2}$ bulk core levels, respectively. While the kagome V$_3$Sn termination shows only the bulk-derived Sn core-level peaks, the ScSn$_2$ termination exhibits additional surface-related components, resulting in four distinct peaks, consistent with previous ARPES studies on ScV$_6$Sn$_6$ \cite{Cheng2024-ch}.

\begin{figure}
	\centering
	\includegraphics{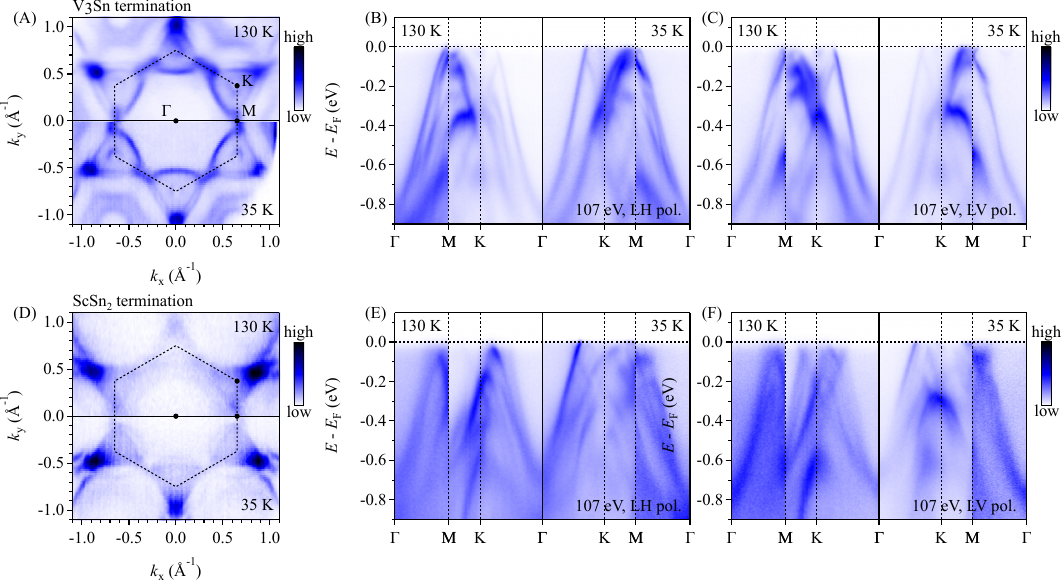}
	\caption{
    ARPES results of ScV$_6$Sn$_6$.
    (A-C) Fermi surface (A) and dispersions along the $\Gamma$-M-K-$\Gamma$ direction (B,C) for the V$_3$Sn termination.
    (D-F) Fermi surface (D) and dispersions along the $\Gamma$-M-K-$\Gamma$ direction (E,F) for the ScSn$_2$ termination.
    The Fermi surface panels show measurements performed above (130 K, top) and below (35 K, bottom) the CDW transition temperature. The left and right sides of panels (B),(C),(E), and (F) correspond to dispersions measured at 130 K and 35 K, respectively. The polarization conditions for the dispersions are indicated in each panel.
    }
	\label{fig:ARPES2}
\end{figure}

The electronic structures of the two terminations are summarized in Fig.~S\ref{fig:ARPES2}. The Fermi surface maps in Figs.~S\ref{fig:ARPES2}A and S\ref{fig:ARPES2}D exhibit clear hexagonal symmetry characteristic of the kagome lattice. However, only the V$_3$Sn termination shows pronounced low-temperature electronic structure renormalization near the M points associated with the CDW transition, which is also evident in the dispersions shown in Figs.~S\ref{fig:ARPES2}B and S\ref{fig:ARPES2}C. The measured dispersions clearly resolve the kagome-derived Dirac cones and van Hove singularities previously reported in ScV$_6$Sn$_6$ \cite{Cheng2024-ch, korshunov2023softening}. Compared to the ScSn$_2$ termination, the V$_3$Sn termination exhibits stronger signatures of CDW-induced band reconstruction, highlighting the importance of the kagome surface in the low-energy electronic instability.

\begin{figure}
	\centering
	\includegraphics{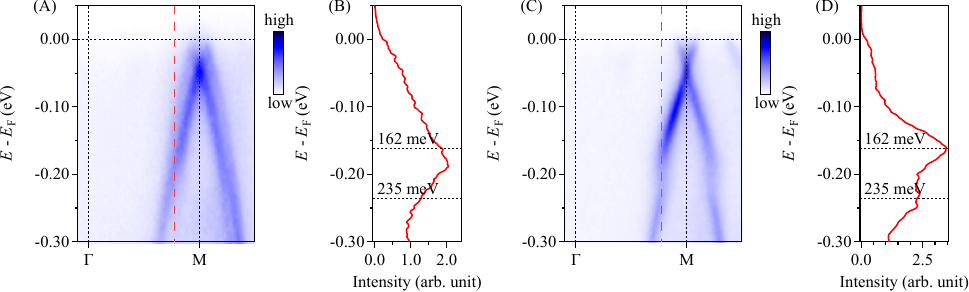}
	\caption{
    CDW gap opening in the $\Gamma$-M dispersion.
    (A) $\Gamma$-M dispersion measured with 107 eV photons under LV polarization at 130 K, above the CDW transition temperature.
    (B) Energy distribution curve (EDC) extracted along the red dashed line in (A).
    (C) $\Gamma$-M dispersion measured with 107 eV photons under LV polarization at 35 K, below the CDW transition temperature.
    (D) EDC extracted along the red dashed line in (C).
    Black dashed lines in (B) and (D) indicate the peak positions associated with the CDW gap opening.
    }
	\label{fig:ARPES3}
\end{figure}

Fig.~S\ref{fig:ARPES3} presents the CDW gap opening along the $\Gamma$-M direction. Compared with the high-temperature dispersions shown in Figs.~S\ref{fig:ARPES3}A and S\ref{fig:ARPES3}B, the low-temperature spectra in Figs.~S\ref{fig:ARPES3}C and S\ref{fig:ARPES3}D exhibit a clear gap opening of approximately $\Delta \sim 75$ meV at $k_{||} = 0.55\mathring{A}^{-1}$ from the $\Gamma$ point along the $\Gamma$-M direction. The corresponding energy distribution curves (EDCs) reveal two split peaks located at 162 meV and 235 meV below the Fermi level. The gap forms near the Brillouin zone boundary associated with the $\sqrt{3}\times\sqrt{3}$ CDW order and is attributed to avoided band crossing caused by CDW-induced band folding. The observed momentum-dependent reconstruction is consistent with strong coupling between the kagome-derived electronic states and the CDW order parameter.

\section{DFT calculations}
Calculations based on density functional theory (DFT)\cite{hohenberg1964inhomogeneous,kohn1965self} were performed using the VASP code \cite{kresse1993ab,kresse1996efficiency, kresse1996efficient}. We calculated the hexagonal structure ($16\times 16\times 8$ k-mesh) and the orthorhombic distortion ($16\times 8\times 8$ k-mesh) with prior relaxation of the unit cell and internal positions and the CDW $\sqrt{3}\times \sqrt{3}\times 3$ reconstruction  reported in \cite{Arachchige_2022} ($9\times 9\times 3$) without relaxation of internal atomic positions. The exchange-correlation potential chosen was the generalized gradient approximation in the Perdew-Burke-Ernzerhof scheme \cite{perdew1996generalized}.  The harmonic phonon spectrum was computed within the finite displacement method implemented in the Phonopy code \cite{phonopy-phono3py-JPCM,phonopy-phono3py-JPSJ}; for all orthorhombic distortions $3\times 2\times 3$ supercells were constructed ($3\times 3\times 2$ k-mesh). The unfolded band structure was post-processed with the VASPKIT package\cite{wang2021vaspkit}.

\begin{figure}[H]
    \centering
    \includegraphics[width=1\linewidth]{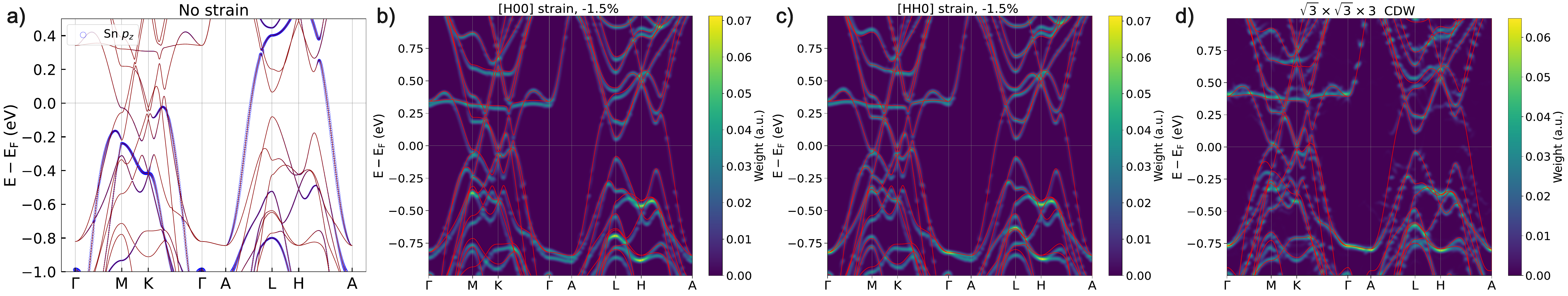}
    \caption{(a) Electronic band structure of ScV$_6$Sn$_6$ in the hexagonal phase with  Sn ${p_z}$  orbital projection. (b,c) Unfolded band structures of the orthorhombic strained structures under compression along the [H00] and [HH0] directions, respectively. d) Unfolded band structure of the  $\sqrt{3}\times \sqrt{3}\times 3$ CDW-reconstructed structure (structure details taken from \cite{Arachchige_2022}). The red solid lines correspond to the normal state (NS) band structure shown in panel (a).}
    \label{fig:supp_bands}
\end{figure}

Figure S\ref{fig:supp_bands}. Panel (a) shows the electronic band structure of ScV$_6$Sn$_6$ in the hexagonal phase with Sn $p_z$ orbital projection. The $p_z$ character is concentrated near the Fermi level along the A-L-H path, reflecting the out-of-plane orbital contribution of the trigonal Sn$^T$ atoms in this energy window. Panels (b) and (c) show the unfolded band structures under 1.5\% compressive strain along the [H00] and [HH0] directions, respectively. In both cases, the unfolded spectral weight closely follows the NS dispersion (red lines), with only minor broadening near selected high-symmetry points and no appreciable energy shifts. Crucially, the $p_z$ spectral weight distribution remains essentially unchanged relative to the unstrained case, indicating that the orthorhombic distortion does not significantly alter the low-energy electronic structure responsible for the lattice instability. Panel (d) shows the CDW-reconstructed  $\sqrt{3}\times \sqrt{3} \times 3$ supercell, where the additional band folding and spectral weight redistribution arise from the enlarged unit cell, but the overall band dispersion of the high-symmetry phase is preserved. The robustness of the electronic structure across all four panels directly supports the interpretation that the strain-induced enhancement of $T_{CDW}$ is not driven by changes in the Fermi surface topology or electron dispersion.

\begin{figure}[H]
    \centering
    \includegraphics[width=1\linewidth]{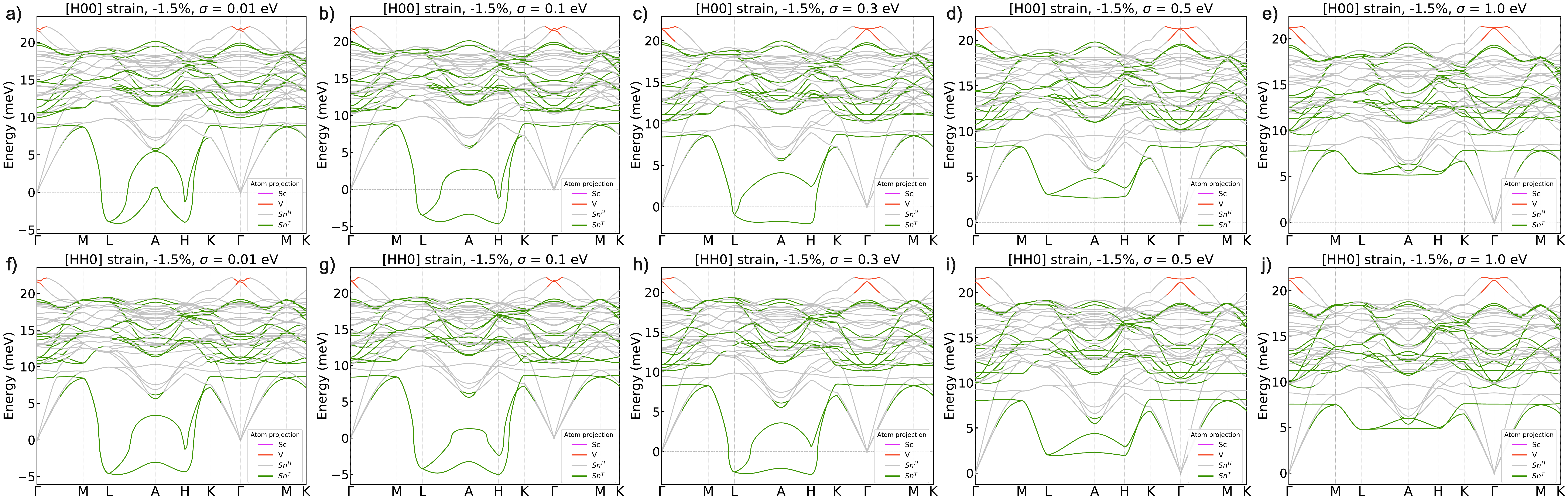}
    \caption{Atomic projection of the phonon dispersion of ScV$_6$Sn$_6$ calculated for increasing Gaussian smearing widths, which mimic increasing electronic temperature. Panels (a–e) correspond to compression along the [H00] direction, while panels (f–j) correspond to compression along the [HH0] direction.}
    \label{supp_electronic_temp}
\end{figure}

Figure S\ref{supp_electronic_temp}. The figure shows the evolution of the phonon dispersion of ScV$_6$Sn$_6$ as a function of Gaussian smearing width, which mimics increasing electronic temperature in the calculations. With increasing smearing, the magnitude of the imaginary phonon frequencies associated with the soft Sn$^T$ branch is progressively reduced, confirming the electronically assisted nature of the CDW instability — the lattice is sensitive to the occupation of states near the Fermi level. Nevertheless, the unstable branch persists over a broad region of reciprocal space even at the largest smearing values, demonstrating that the underlying lattice instability is robust and not purely electronically driven. This behavior is qualitatively identical for compression along both the [H00] and [HH0] directions, indicating that the thermal suppression of the instability has no significant dependence on the in-plane orientation of the orthorhombic distortion.

\begin{figure}[H]
    \centering
    \includegraphics[width=1\linewidth]{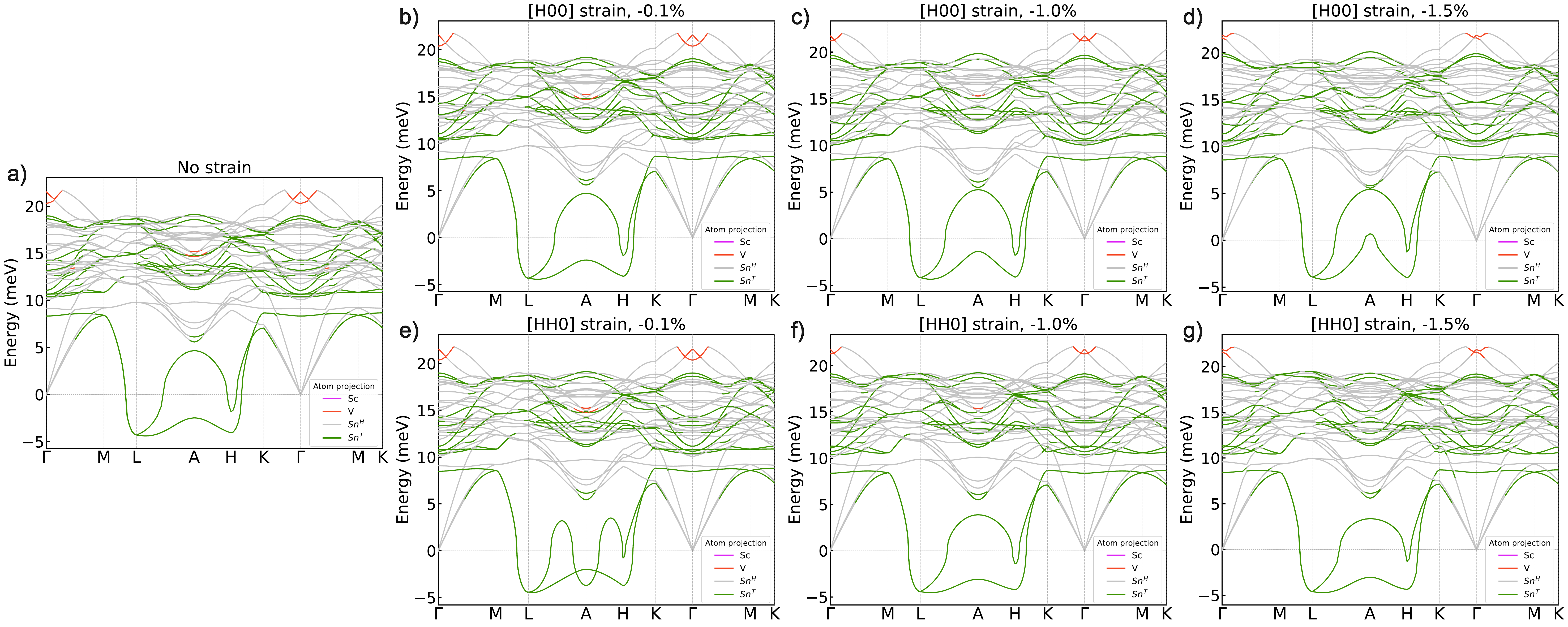}
    \caption{Atomic projection of the phonon dispersion of ScV$_6$Sn$_6$ for (a) the unstrained structure and under increasing compressive strain along the (b–d) [H00] and (e–g) [HH0] directions. }
    \label{fig:supp_strain}
\end{figure}

Figure S\ref{fig:supp_strain}. Panel (a) shows the reference phonon dispersion of the unstrained structure, with the characteristic nearly flat imaginary branch along the L–A–H path dominated by out-of-plane Sn$^T$ vibrations. Panels (b–d) and (e–g) show the evolution under increasing compression along [H00] and [HH0], respectively. In both cases, the imaginary branch persists across the majority of the Brillouin zone at all strain levels, demonstrating that orthorhombic distortion does not destroy the underlying lattice instability. A partial stabilization near the A point develops under the largest applied compression, and this effect appears marginally more pronounced along [H00] than [HH0], consistent with [H00] corresponding to compression along the longer orthorhombic axis. Beyond this subtle difference, the phonon spectra under the two strain directions are qualitatively similar throughout, with no strongly anisotropic modification of the soft branch. This weak directional dependence rules out a scenario in which the enhancement of $T_{CDW}$ is governed by anisotropic changes in the electron–phonon coupling, and instead points to the strain-driven ordering of the frustrated Sn$^T$–Sc–Sn$^T$ trimer configurations as the dominant mechanism, as discussed in the main text.

\section{Monte Carlo Simulations}
\subsection{Simulation Setup -- Unstrained}

\begin{figure}
	\centering
	\includegraphics[width=0.8\linewidth]{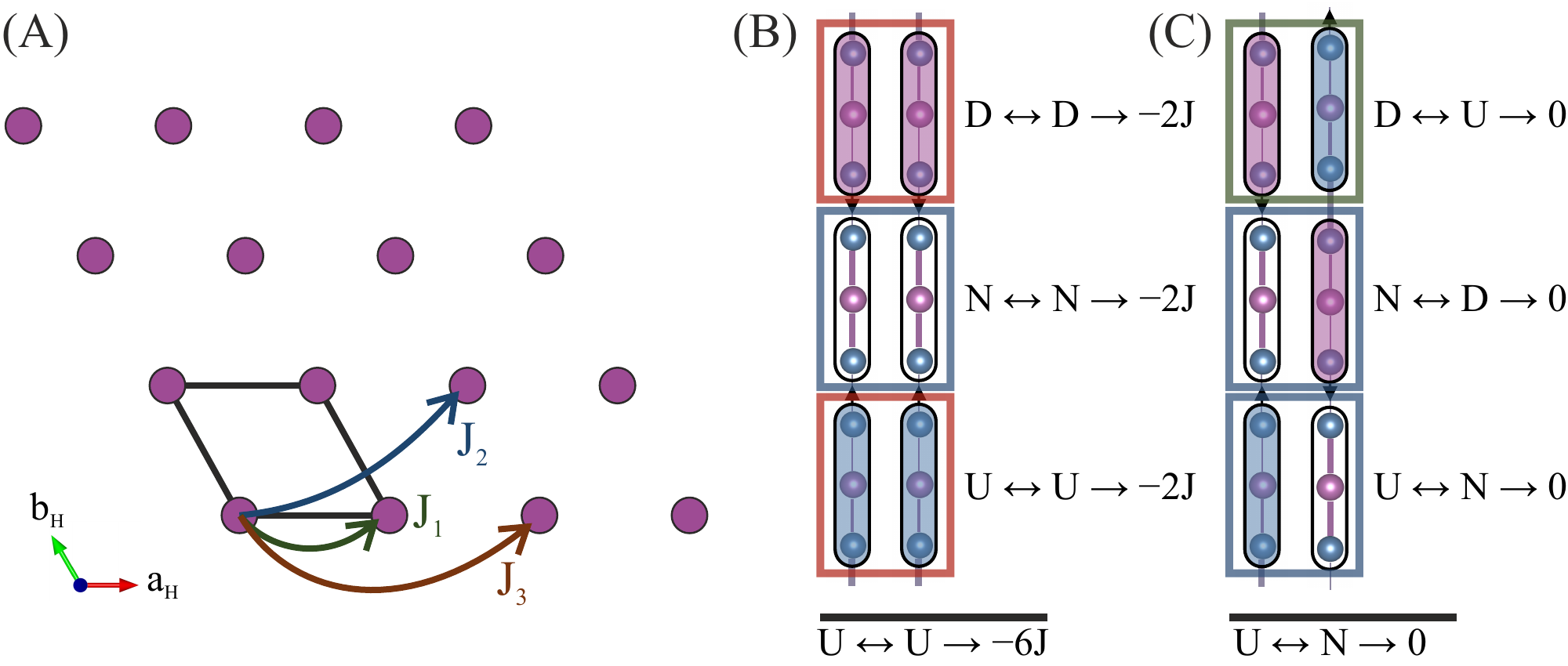}
	\caption{ (A) Illustration of the exchange interaction parameters on the hexagonal lattice. (B-C) Illustration of the effective exchange Hamiltonian. (B) Neighbouring ``UU'' pseudo-atom configuration. For each UU or DD pair within the chain of three trimers, an energy penalty of $-J_1$ is assigned, yielding a total penalty of $-2J_1$ for the trimer-chain configuration. (C) Neighbouring ``UN'' pseudo-atom configuration. For each UD pair within the chain of three trimers, an energy gain of $+J_1$ is assigned, yielding a total energy gain of $+J_1$ for the trimer-chain configuration.}
	\label{fig:SIMC_J}
\end{figure}

Monte Carlo (MC) simulations were performed on a two-dimensional supercell of $90\times90\times1$ unit cells using a coarse-grained approach, in which Sc--Sn--Sc trimers were assigned one of three configurations: up (U), down (D), or neutral (N). 
All simulations employed a hexagonal geometry, with the lattice initially populated with equal probabilities for each configuration. The simulations were implemented in a custom Fortran script. We assume the system to be fully ordered along the $c$-direction, as illustrated in Figure~1B of the main text. The pseudo-atoms used in the MC simulation are labelled according to the atom at the bottom layer of the configuration shown in Figure~1B of the main text (see also Figure~S\ref{fig:SIMC_J}).

For the unstrained simulations, interaction parameters reported by \citet{Gomez_2024} were mapped from the two-component Ising system to a three-state Potts system, as illustrated in Figure~S\ref{fig:SIMC_J} ($J_1 = -6.56$~K, $J_2 = +1.74$~K, and $J_3 = -1.03$~K). Same likewise neighbours receive double the penalty as compared to the Ising model, while no energy gain is implemented for un-likewise neighbours.

To prevent the system from becoming trapped in local energy minima, simulated annealing was employed. The annealing procedure consisted of 20 temperature steps, with a starting temperature defined as $T_\mathrm{start} = \max(100~\mathrm{K},\, 3 \times T_\mathrm{final})$. At each step, a number of MC moves equal to 50 times the total number of atoms was attempted. 
Each MC move consisted of swapping two pseudo-atoms in different states. A move was accepted if a uniformly distributed random number in $[0,1]$ was less than $\exp\!\left(-\Delta E / T_\mathrm{MC}\right)$, where $\Delta E$ is the energy 
difference between the proposed and current configurations.

A pseudo-Ising Hamiltonian was used, in which the pairwise correlation parameters derived by \citet{Gomez_2024} are mapped onto the trimer pseudo-atoms, as illustrated in Figure~S\ref{fig:SIMC_J}(B, C). Like nearest-neighbour pairs incur an energy penalty of 
$-2J$, while unlike pairs yield an energy gain of $+J$. Note the sign convention of $J_1$, which determines whether nearest-neighbour interactions result in an energy gain or penalty.

The MC configurations were expanded to $90\times90\times3$ unit cells by explicitly placing the Sc and Sn atomic coordinates corresponding to each up, down, or neutral trimer. Single-crystal diffuse scattering was then calculated from these configurations using the DISCUS program~\cite{neder2008diffuse}. The diffuse scattering was computed on a grid of $181\times181\times31$ voxels covering $-1 \leq h, k \leq 1$ and $-5 \leq l \leq 5$, with step sizes $\Delta h = \Delta k = 1/90$ and $\Delta l = 1/3$, using the NUFFT Fourier algorithm as implemented in DISCUS. The diffuse scattering from five independent configurations was averaged and symmetrized according to $2/m$ Laue symmetry with the unique axis along $c$, in order to not suppress any  local orthorhombic distortions in the diffuse scattering signal.

\begin{figure}
	\centering
	\includegraphics[width=.7\linewidth]{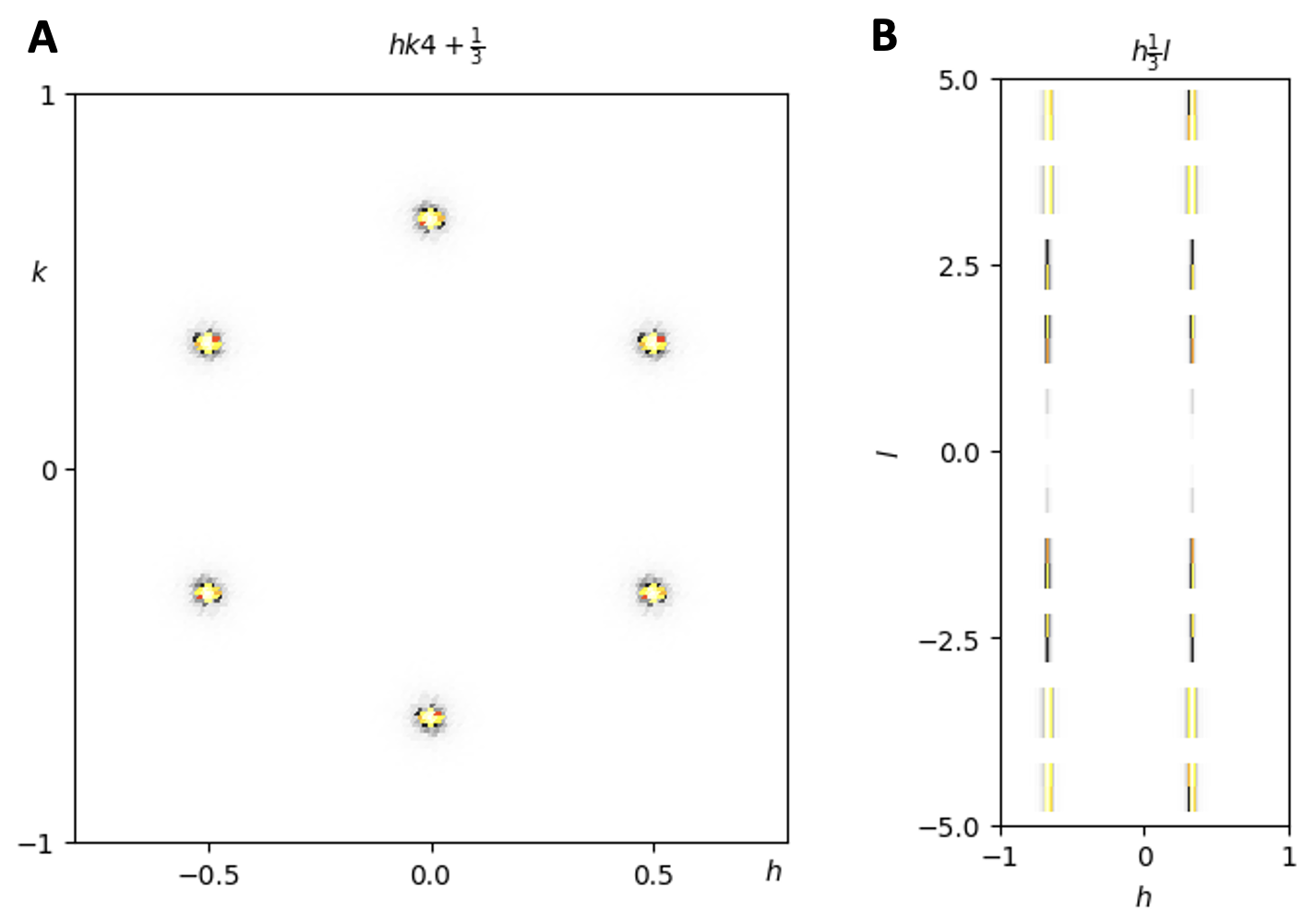}
	\caption{Calculated diffuse scattering for zero strain and $T_\mathrm{MC} = 0$~K. 
		(A) $hk4.33$ layer. 
		(B) $h\frac{1}{3}l$ layer. Note that the limited resolution along $l$ is due to the coarse simulation grid along $c$.}
	\label{fig:SIMC_Diffuse}
\end{figure}

The resulting diffuse scattering in the $hk4.33$ layer for $T_\mathrm{MC} = 0$~K is shown in Figure~S\ref{fig:SIMC_Diffuse}, in agreement with \cite{Gomez_2024}.

\subsection{Temperature Dependence}

MC simulations were performed for $T_\mathrm{MC}$ values ranging from 0 to 300~K in steps of 5~K. To estimate the CDW transition temperature, the calculated diffuse scattering intensity in the vicinity of the $\left(\frac{1}{3},\frac{1}{3},4+\frac{1}{3}\right)$ satellite reflection integrated in the hk$4+\frac{1}{3}$-layer, with an integration radius of 9 pixels, which equals $\Delta h = 0.1$, to yield the satellite intensity $I_\mathrm{sat}$.

\begin{figure}
	\centering
	\includegraphics[width=.7\linewidth]{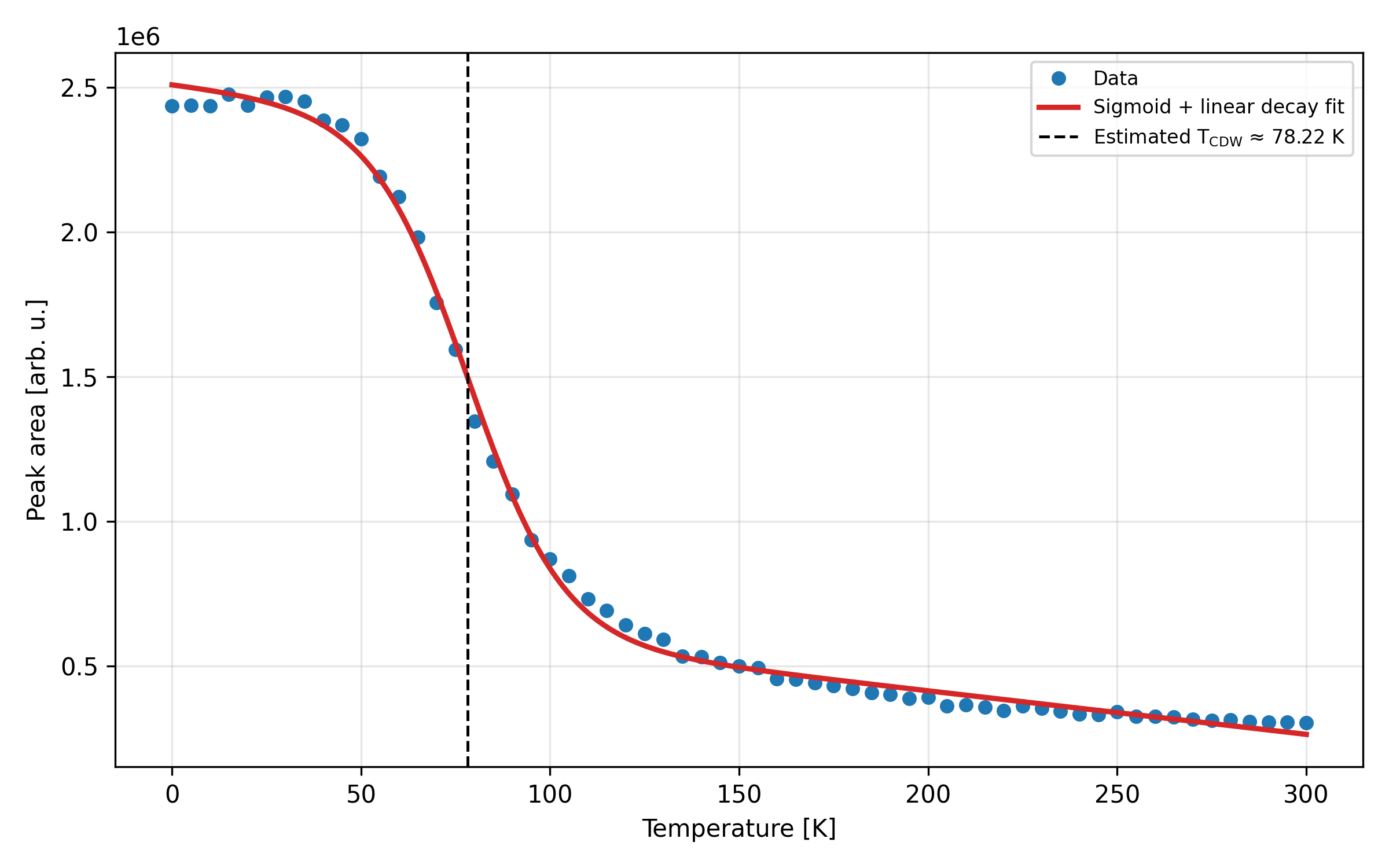}
	\caption{Fit of the integrated satellite peak area as a function of Monte Carlo Temperature with a Sigmoid function with linear decay for 0~\% strain.}
	\label{fig:SIMC_Sigmoid}
\end{figure}

The CDW transition temperature $T_\mathrm{CDW}$ was then estimated by fitting a sigmoid function with a linear decay to $I_\mathrm{sat}$ as a function of $T_\mathrm{MC}$:
\begin{equation}
	I_\mathrm{sat}(T) = I_\mathrm{low} + 
	\frac{I_\mathrm{high} - I_\mathrm{low}}
	{1 + \exp\!\left(-\dfrac{T - T_\mathrm{CDW}}{\delta}\right)} + \alpha (T_{150 \mathrm{K}}),
\end{equation}
where $I_\mathrm{low}$, $I_\mathrm{high}$, $T_\mathrm{CDW}$, the transition width $\delta$  and the slope $\alpha$ are free fitting parameters. Figure~S\ref{fig:SIMC_Sigmoid} shows an example fit.

In a similar fashion the CDW transition temperature $T_\mathrm{CDW}$ can be estimated from the final MC Energy and the Warren-Cowley short range order parameter $\alpha$ for the first neighbour U-D pairs. 

\subsection{Strain integration}
To integrate the strain into the MC simulations, the interaction parameters for the nearest-neighbours were split into symmetry-independent pairs (compare Figure 4 of the main text).
The interaction parameters were adapted depending on the new interatomic distance according to:

\begin{equation}
    J_i(d) = J_{i0} \times \exp\left[\lambda \left(1 - \frac{d}{d_0}\right)\right]
\end{equation}  

where $J_0$ is the interaction parameter for the unstrained system and $d_0$ is the interatomic distance in the unstrained system. The parameter \(\lambda\) controls the sensitivity of the effective coupling constants to changes in the interatomic distances. Its value was chosen by performing MC simulations for several \(\lambda\) values and selecting \(\lambda=25\), which reproduces the experimentally observed slope of \(T_{\mathrm{CDW}}\) as a function of applied strain.

For the lattice compression along [H00] direction, the lattice parameters change as $a_{orth}=\sqrt{3} x b_{hex} $ and $b_{orth} = b_{hex}$, where $x\le 1$ is the compression factor. The resulting interatomic distances for the considered neighbour pairs:

\begin{equation}
	\begin{aligned}
	d_{j_{1a}} = \frac{1}{2} \sqrt{a_{orth}^2 + b_{orth}^2} = \frac{1}{2} \sqrt{3x^2+1}b_{hex} \\
	d_{j_{1b}} =b_{hex} \\
	d_{j_{2a}} = a_{orth}=  \sqrt{3}xb_{hex} \\
	d_{j_{2b}} = \frac{1}{2} \sqrt{a_{orth}^2 + 9b_{orth}^2} = \frac{\sqrt{3}}{2} \sqrt{x^2+3}b_{hex} \\
	d_{j_{3a}} =  \sqrt{a_{orth}^2 + b_{orth}^2} =  \sqrt{3x^2+1}b_{hex} \\
	d_{j_{3b}} =2b_{hex} \\
	\end{aligned}
\end{equation}

In contrast, the compression along the [HH0] direction is equivalent to the compression along the short orthorhombic axis, with $a_{orth}=\sqrt{3} b_{hex} $ and $b_{orth} = x b_{hex}$. The resulting changes in the interatomic distances can be written as:
\begin{equation}
	\begin{aligned}
	d_{j_{1a}} = \frac{1}{2} \sqrt{a_{orth}^2 + b_{orth}^2} = \frac{1}{2} \sqrt{3+x^2}b_{hex} \\
	d_{j_{1b}} =xb_{hex} \\
	d_{j_{2a}} = a_{orth}=  \sqrt{3}b_{hex} \\
	d_{j_{2b}} = \frac{1}{2} \sqrt{a_{orth}^2 + 9b_{orth}^2} = \frac{\sqrt{3}}{2} \sqrt{1+3x^2}b_{hex} \\
	d_{j_{3a}} =  \sqrt{a_{orth}^2 + b_{orth}^2} =  \sqrt{3+x^2}b_{hex} \\
	d_{j_{3b}} =2xb_{hex} \\
	\end{aligned}
\end{equation}

\begin{figure}
	\centering
	\includegraphics[width=.7\linewidth]{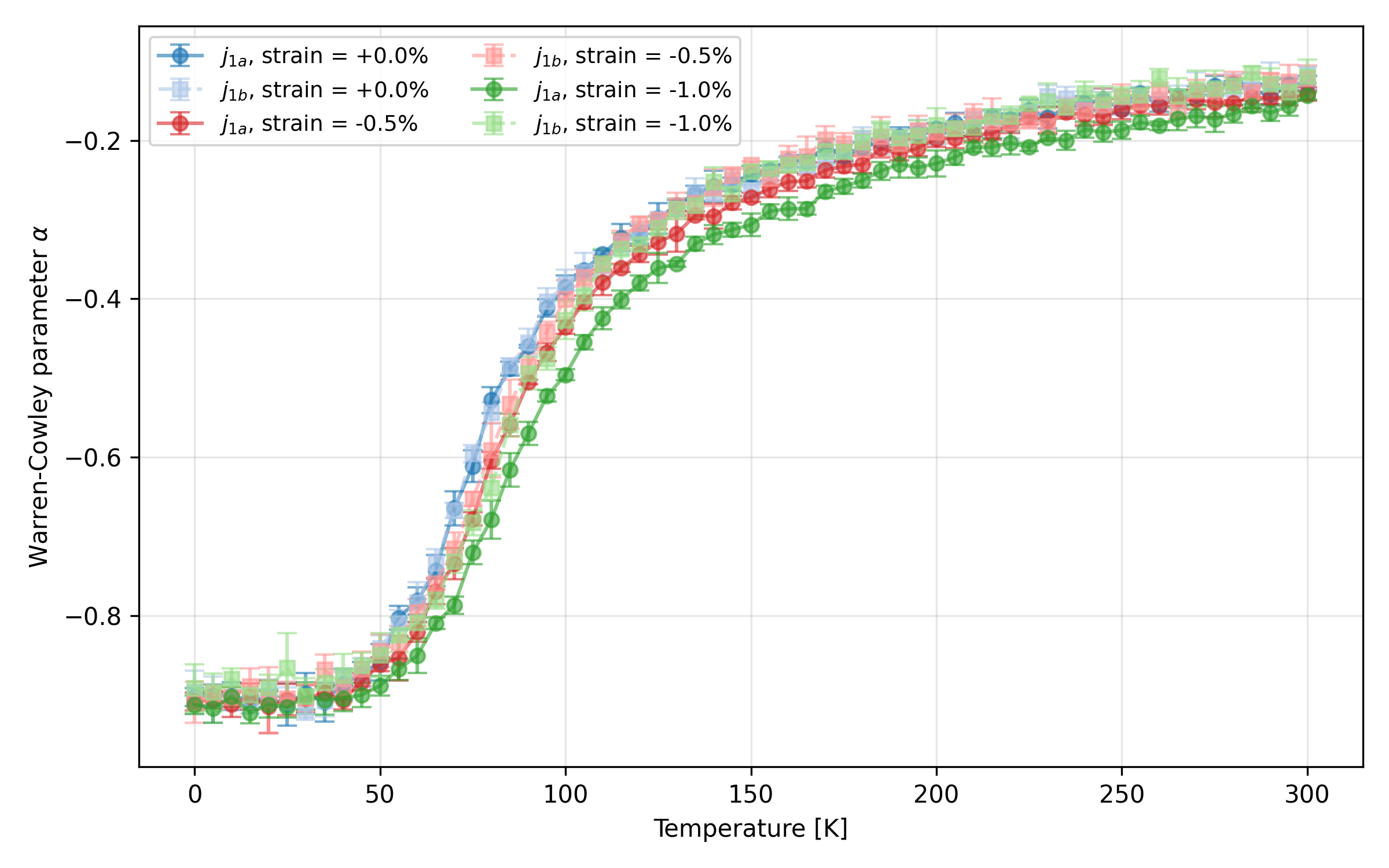}
	\caption{Warren-Cowley short range order parameter $\alpha$ for the U-D pairs for interaction $J_{1a}$ and $J_{1b}$ for three selected strains.}
	\label{fig:SIMC_alpha_split}
\end{figure}

Figure~S\ref{fig:SIMC_alpha_split} shows the Warren-Cowley short range order parameter for U-D pairs in the unstrained, 0.5~\% and 1~\% strained structures.

\begin{figure}
	\centering
	\includegraphics[width=.7\linewidth]{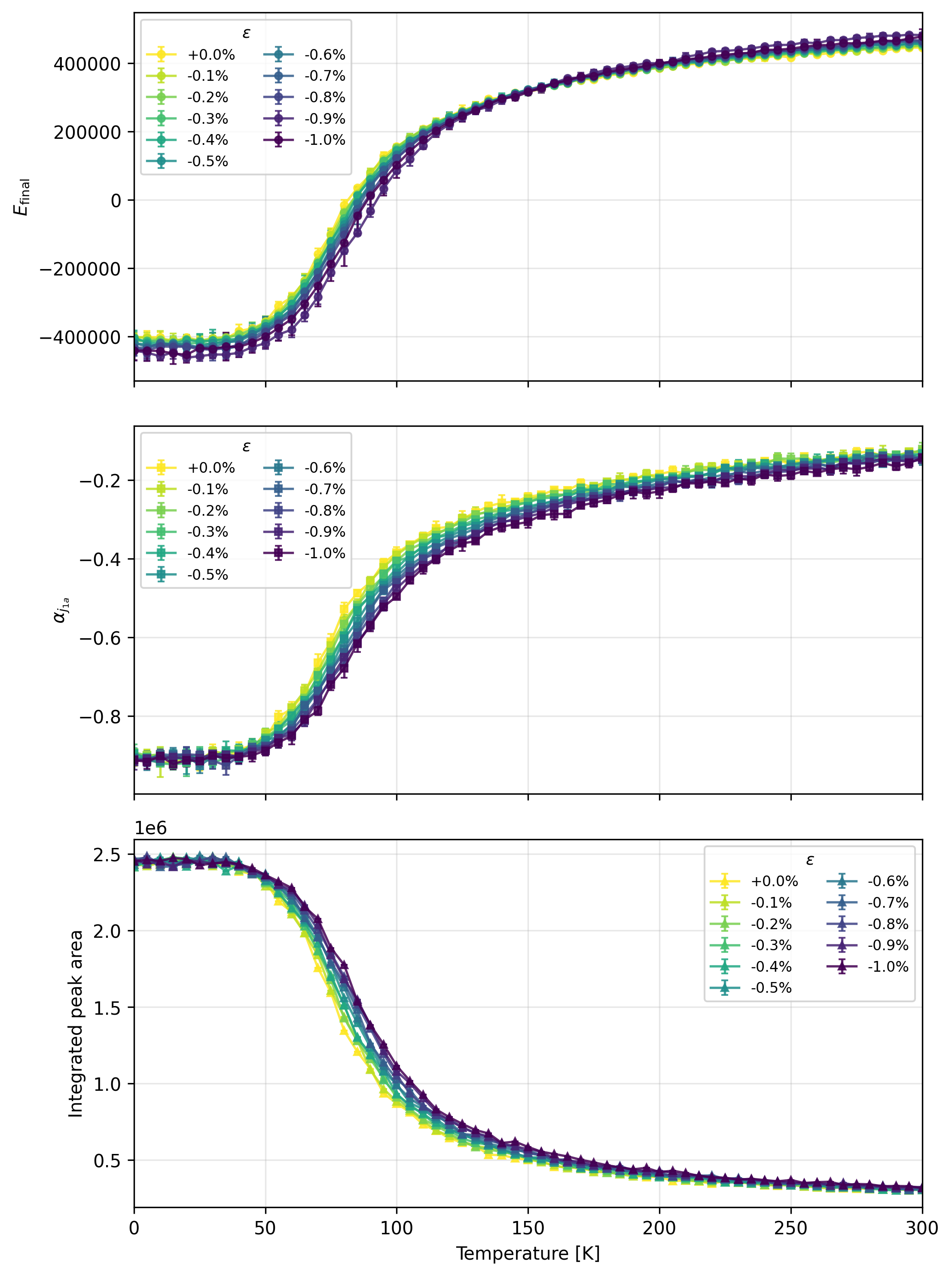}
	\caption{Final MC energy, Warren-Cowley short range order parameter resulting from the $J_{1a}$ interaction, and integrated peak area as a function of the simulation temperature and strain.}
	\label{fig:SIMC_threePanels}
\end{figure}

\begin{figure}
	\centering
	\includegraphics[width=1.0\linewidth]{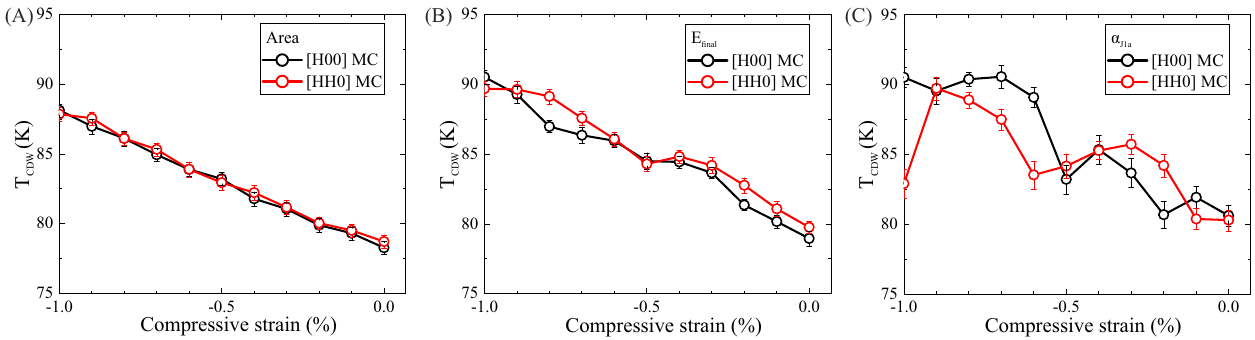}
	\caption{Fitted $\mathrm{T}_\mathrm{CDW}$ as a function of strain for the [H00] and [HH0] directions, extracted from the integrated peak area (A), the final MC energy (B),  and the Warren-Cowley short range order parameter resulting from the $J_{1a}$ interaction (C).}
	\label{fig:SIMC_TCDW}
\end{figure}

Figure~S\ref{fig:SIMC_threePanels} shows the progression of the final MC energy, the integrated peak intensity and the Warren-Cowley short range order parameter $\alpha$ for the first neighbour U-D pairs as a function of strain and simulation temperature. The fitted transition temperatures are shown in Figure~S\ref{fig:SIMC_TCDW}.

\newpage
\bibliography{biblio}